\documentclass[prb, aps, 10pt, twocolumn, floatfix, superscriptaddress]{revtex4-2}
\usepackage{amssymb,amsfonts,amsmath,amsthm}
\usepackage{graphicx}
\usepackage[english]{babel}
\usepackage[latin1]{inputenc}
\usepackage[pdftex,colorlinks=true,pdfstartview=FitH,linkcolor=blue,citecolor=blue,urlcolor=blue]{hyperref}

\usepackage{bm}
\usepackage{float}
\usepackage{mathtools}
\usepackage{color,soul}
\usepackage{times}
\usepackage{upgreek}
\usepackage{MnSymbol}
\usepackage{verbatim}
\usepackage[bottom]{footmisc}
\usepackage{bbold}

\usepackage[dvipsnames,svgnames,table]{xcolor}
\usepackage{hyperref}
\usepackage{fancyhdr}
\usepackage[english]{babel}
\usepackage[caption=false]{subfig}
\usepackage{orcidlink}

\usepackage[scr=rsfs,cal=boondox]{mathalfa}

\hypersetup{
pdfstartview={FitH},
colorlinks=true,    
linkcolor=NavyBlue, 
citecolor=Blue,   
filecolor=NavyBlue, 
urlcolor=NavyBlue   
}

\newcommand{\bea}{\begin{eqnarray}}
\newcommand{\eea}{\end{eqnarray}}
\newcommand{\be}{\begin{equation}}
\newcommand{\ee}{\end{equation}}

\begin{document}
\title{Theory of optomechanical locking in driven-dissipative coupled polariton condensates}

\author{I. A. Ramos-P\'erez\orcidlink{0000-0001-9796-8461}}
\thanks{These three authors contributed equally}
\affiliation{Centro At{\'{o}}mico Bariloche and Instituto Balseiro,
Comisi\'on Nacional de Energ\'{\i}a At\'omica (CNEA)- Universidad Nacional de Cuyo (UNCUYO), 8400 Bariloche, Argentina.}
\affiliation{Instituto de Nanociencia y Nanotecnolog\'{i}a (INN-Bariloche), Consejo Nacional de Investigaciones Cient\'{\i}ficas y T\'ecnicas (CONICET), Argentina.}

\author{I. Carraro-Haddad\orcidlink{0009-0002-1635-2732}}
\thanks{These three authors contributed equally}
\affiliation{Centro At{\'{o}}mico Bariloche and Instituto Balseiro,
Comisi\'on Nacional de Energ\'{\i}a At\'omica (CNEA)- Universidad Nacional de Cuyo (UNCUYO), 8400 Bariloche, Argentina.}
\affiliation{Instituto de Nanociencia y Nanotecnolog\'{i}a (INN-Bariloche), Consejo Nacional de Investigaciones Cient\'{\i}ficas y T\'ecnicas (CONICET), Argentina.}

\author{F. Fainstein\orcidlink{0000-0002-3700-0106}}
\thanks{These three authors contributed equally}
\affiliation{Universidad de Buenos Aires, Facultad de Ciencias Exactas y Naturales, Departamento de F\'{i}sica, Ciudad Universitaria, 1428 Buenos Aires, Argentina.}
\affiliation{CONICET - Universidad de Buenos Aires, Instituto de F\'{i}sica Interdisciplinaria y Aplicada (INFINA), Ciudad Universitaria, 1428 Buenos Aires, Argentina.}

\author{D. L. Chafatinos\orcidlink{0000-0001-8481-8958}}
\affiliation{Centro At{\'{o}}mico Bariloche and Instituto Balseiro,
Comisi\'on Nacional de Energ\'{\i}a At\'omica (CNEA)- Universidad Nacional de Cuyo (UNCUYO), 8400 Bariloche, Argentina.}
\affiliation{Instituto de Nanociencia y Nanotecnolog\'{i}a (INN-Bariloche), Consejo Nacional de Investigaciones Cient\'{\i}ficas y T\'ecnicas (CONICET), Argentina.}

\author{G. Usaj\orcidlink{0000-0002-3044-5778}}
\affiliation{Centro At{\'{o}}mico Bariloche and Instituto Balseiro,
Comisi\'on Nacional de Energ\'{\i}a At\'omica (CNEA)- Universidad Nacional de Cuyo (UNCUYO), 8400 Bariloche, Argentina.}
\affiliation{Instituto de Nanociencia y Nanotecnolog\'{i}a (INN-Bariloche), Consejo Nacional de Investigaciones Cient\'{\i}ficas y T\'ecnicas (CONICET), Argentina.}
\affiliation{TQC, Universiteit Antwerpen, Universiteitsplein 1, B-2610 Antwerpen, Belgium}
\affiliation{CENOLI, Universit\'e Libre de Bruxelles - CP 231, Campus Plaine, B-1050 Brussels, Belgium}

\author{G.~B. Mindlin\orcidlink{0000-0002-7808-5708}}
\affiliation{Universidad de Buenos Aires, Facultad de Ciencias Exactas y Naturales, Departamento de F\'{i}sica, Ciudad Universitaria, 1428 Buenos Aires, Argentina.}
\affiliation{CONICET - Universidad de Buenos Aires, Instituto de F\'{i}sica Interdisciplinaria y Aplicada (INFINA), Ciudad Universitaria, 1428 Buenos Aires, Argentina.}

\author{A. Fainstein\orcidlink{0000-0001-7875-3609}}
\affiliation{Centro At{\'{o}}mico Bariloche and Instituto Balseiro,
Comisi\'on Nacional de Energ\'{\i}a At\'omica (CNEA)- Universidad Nacional de Cuyo (UNCUYO), 8400 Bariloche, Argentina.}
\affiliation{Instituto de Nanociencia y Nanotecnolog\'{i}a (INN-Bariloche), Consejo Nacional de Investigaciones Cient\'{\i}ficas y T\'ecnicas (CONICET), Argentina.}

\author{A.~A. Reynoso\orcidlink{0000-0002-5457-0020}}
\email[Corresponding author, e-mail: ]{reynoso@cab.cnea.gov.ar}
\affiliation{Centro At{\'{o}}mico Bariloche and Instituto Balseiro,
Comisi\'on Nacional de Energ\'{\i}a At\'omica (CNEA)- Universidad Nacional de Cuyo (UNCUYO), 8400 Bariloche, Argentina.}
\affiliation{Instituto de Nanociencia y Nanotecnolog\'{i}a (INN-Bariloche), Consejo Nacional de Investigaciones Cient\'{\i}ficas y T\'ecnicas (CONICET), Argentina.}
\affiliation{Departamento de F\'isica Aplicada II, Universidad de Sevilla, E-41012 Sevilla, Spain.}

\date{\today}

\begin{abstract}
{The ubiquitous phenomenon of synchronization is inherently characteristic of dynamical dissipative non-linear systems. In particular, synchronization has been theoretically and experimentally demonstrated for exciton-polariton condensates with time-independent coupling. Unlike that case, we theoretically investigate the effects of a fixed mechanical harmonic driving, i.e., a coherent phonon population, that induces time modulation in the coupling of two coupled condensates. Our model applies both to electrically generated modulation, through coherent bulk acoustic waves, and to self-induced optomechanical vibrations. We consider linear as well as quadratic phonon-displacement-induced modulations of the polariton coupling. Peculiar asynchronously locked phases are found and analyzed in the context of synchronization and Josephson-type oscillations phenomena that appear in the non-driven case. Notably, Arnold tongues corresponding to the asynchronously locked phases emerge at condensate detunings that correspond to integer numbers of the mechanical frequency and also to rational fractions of it. Unlocked quasiperiodic and chaotic regimes can also be reached. In particular, the phonon-induced fractional locking frequencies arrange in a Farey sequence that produces a devil's staircase of the steady-state dressed detuning between the condensates. Importantly, both polariton-polariton and reservoir-polariton interactions facilitate the realization of coherent phonon-induced asynchronously locked phases. The relation to recent experiments on optomechanically driven condensates in arrays of polariton traps is discussed. }
\end{abstract}
\maketitle
\section{Introduction}

Microcavity exciton-polariton fluids, a quantum state of matter formed by strongly coupled excitons and
photons in microcavities made from semiconductors, constitute a hybrid system~\cite{Kurizki2015}  that displays a plethora of interesting properties. These include, among others, Bose-Einstein condensation~\cite{Kasprzak2006}, superfluidity~\cite{Amo2009}, and Josephson-type oscillations~\cite{Lagoudakis2010,Abbarchi2013}. While these fluids bear some similarities with more standard quantum phases in interacting equilibrium systems, there are several peculiar aspects that arise from their driven-dissipative non-equilibrium nature that makes them especially interesting, even more nowadays with well-established ideas on the field of non-Hermitian dynamics~\cite{CarusottoRMP2013}. The recent blooming of polariton related research has been strongly supported on the degree of maturity that the experimental capability of engineering controllable coupled polariton traps has attained, either in pairs~\cite{Lagoudakis2010,Abbarchi2013}  or forming arrays of different geometries and dimensionalities~\cite{Hartmann2006,Winkler2015,Kuznetsov2018,Alyatkin2020,Kalinin2020}. This has allowed, for instance, the implementation of quantum simulators~\cite{Kim2017,Kalinin2020b,Boulier2020,Gosh2020} and the exploration of different topological properties of known lattice models~\cite{Solnyshkov2021}, profiting from the experimental possibility of accessing excited states and/or controlling interactions among polaritons.

In parallel, another emerging area is that of cavity optomechanics, hybrid structures that bridge the optical domain with acoustics~\cite{RMP,Thomas2006,PainterOMX,RMP,Ren2020}. Optomechanical resonators and optomechanical crystals exploit the colocalization of mechanical and optical modes to greatly enhance their interaction. Non-linearities arising in this context due to dynamical feedback lead, e.g., to the optical cooling of the mechanical oscillator down to its quantum ground state~\cite{O'Connell,Teufel,Chan,Verhagen}, and also to mechanical self-oscillation, conceptually similar to phonon lasing~\cite{Kippenberg,Grudinin}. The field of cavity optomechanics has recently intersected with that of exciton-polariton fluids in semiconductor microcavities due to the serendipitous discovery that the same GaAs/AlAs-based distributed Bragg reflector planar microcavities supporting cavity polaritons also confine GHz hypersound~\cite{Fainstein2013,Kimura2015}. This is particularly relevant in view of the potential access to a very strong optomechanical coupling mediated by excitons and based on the deformation potential interaction~\cite{Jusserand2015,Scherbakov2022}. In fact, it has been recently shown that GHz coherent mechanical waves that strongly modulate the polariton states can be either self-induced by continuous-wave optical excitation~\cite{Chafatinos2020,Reynoso2022,Chafatinos2022,Kuznetsov2023}, or externally injected using piezoelectric bulk acoustic wave (BAW) transducers~\cite{Kuznetsov2023,Kuznetsov2021,Crespo-Poveda2022}. These resources expand the cavity-optomechanics toolkit that, for example, could be used to implement effective gauge potentials in polaritonics based on the mechanically induced dynamic modulation of inter-site coupling in lattices, as originally proposed to demonstrate the photonic Aharonov-Bohm effect in photonic fibers~\cite{Fang2012}, and more recently non-Hermitian chiral phononics in photonic crystal nano-beams~\cite{DelPino2022}.

Understanding in detail how phonons reshape light-matter coupled states is a question that naturally arises at the frontier of polaritonics and cavity optomechanics. The interplay between driven-dissipative dynamics and interactions, which sets exciton-polaritons apart from other quantum fluids, is expected to play a significant role in the response to coherent phonons. Does the modulated coupling between condensates merely introduce time dependence and a phase shift, or does it qualitatively alter these states? To address this question we theoretically investigate a system of two neighboring coupled condensates driven above the condensation threshold with a continuous, non-resonant laser. Here, we demonstrate striking signatures of coherent phonon-induced modulation of the coupling. Remarkably, the states in adjacent traps can lock asynchronously, exhibiting energy differences that are integer multiples, or even rational fractions, of the confined phonon energy. This phenomenon, which is shown to be enhanced by interactions, paves the way for phonon-mediated control of non-resonantly driven condensate dynamics. Importantly, our findings may also extend to electrically driven quantum fluids, as demonstrated in recent experiments on electrically driven non-resonant systems~\cite{Schneider2013}.

In the following sections, we deal with a model of two trapped coupled polariton condensates that explicitly includes the interplay between driven-dissipative dynamics and interactions. Our approach builds upon the rich body of work on synchronization and Josephson oscillations in non-Hermitian coupled polariton condensates for time-independent inter-trap coupling ~\cite{Wouters2008, Eastham2008, Eastham2021, Baas2008, Ohadi2018}. Before including the effect of a coherent population of GHz phonons, we begin our analysis by discussing such time-independent inter-trap coupling case from a dynamical systems perspective. We then progressively incorporate the effect of the time-periodic mechanical perturbation, first in a rotating-wave approximation, and then through a full numerical modeling of the problem. We focus on two experimentally relevant cases in which the mechanics modulates either linearly or quadratically the inter-trap coupling. We analyze the resulting dynamics and identify the conditions under which the above-mentioned asynchronously locked polariton states are stable. Finally, we conclude by presenting a detailed account of our findings and their implications in potential experimental realizations.


\section{The physical system}

\begin{figure}[!hht]
 \begin{center}
    \includegraphics[trim = 10mm 0mm 0mm -10mm,clip=true, keepaspectratio=true, width=0.99\columnwidth]{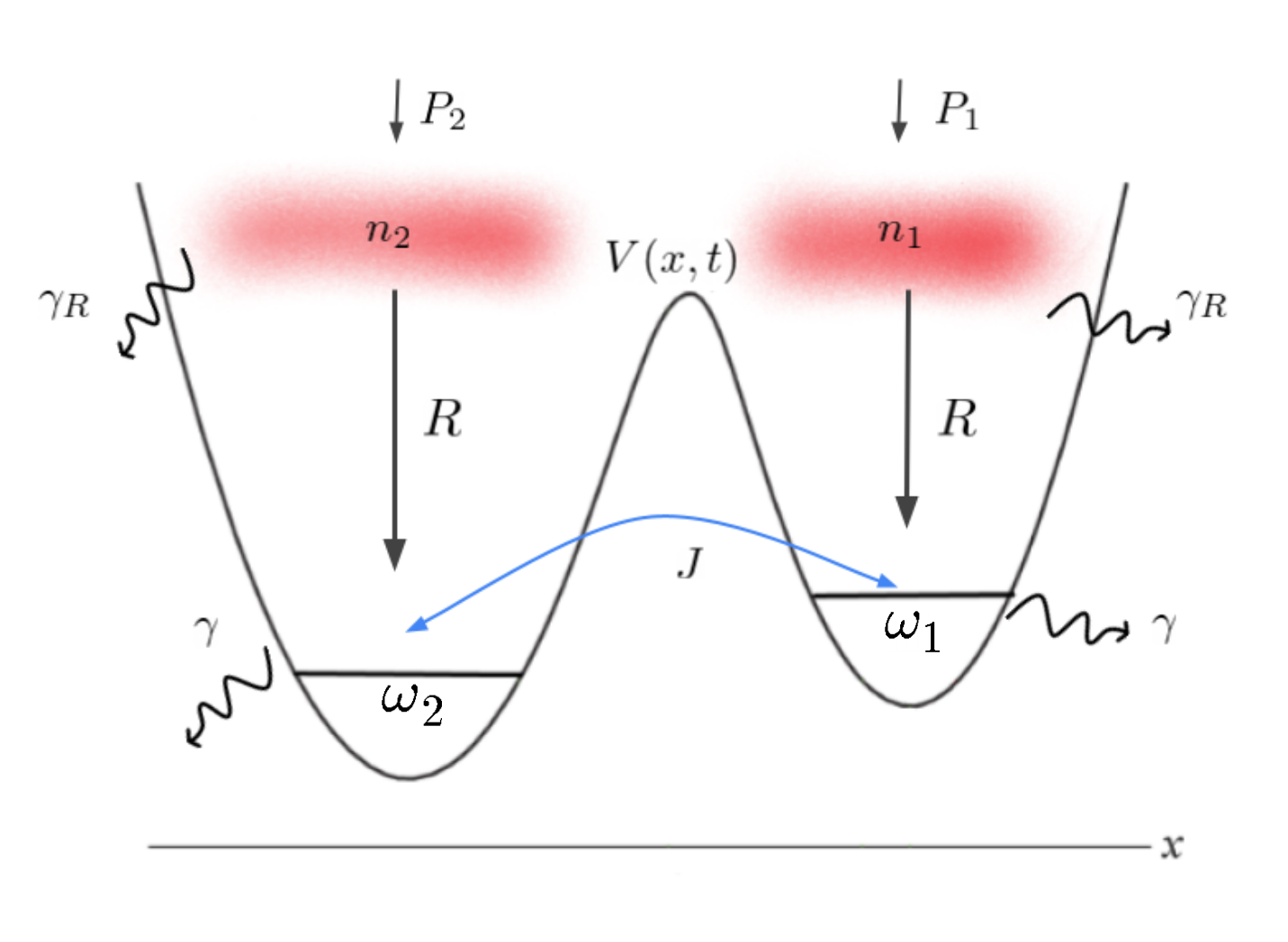}
\end{center}
\vspace{-1.3 cm}
\hspace{0 cm}
\caption{
Scheme of polariton condensates in a pair of coupled traps. $\omega_j$ is the bare frequency of the $j$-mode (for clarity in the scheme a single confined state is assumed per trap), $J$ describes a direct hopping rate term between modes [determined by the potential $V(x,t)$], $\gamma$ is the polariton decay rate and $R$ the stimulated loading rate from the reservoir. The dynamic of the latter is controlled by the pump power $P_j$, the excitonic decay rate $\gamma_\mathrm{R}$ and the stimulated decay to the condensate. The time dependence in the potential $V(x,t)$ arises from the modulation introduced by a coherent mechanical wave, and which can affect the hopping rate $J$.
Polaritons are affected by Coulomb interactions through $U_j^{\textrm{P}}$ and $U_j^{\textrm{R}}$, the polariton-polariton and reservoir-polariton couplings, respectively.
}
\label{Fig1}
\end{figure}

The physical system considered consists on pairs of coupled $\mathrm{\mu}$m-sized traps based on distributed Bragg reflector (DBR) cavities as those previously studied in the context of polariton phenomena~\cite{CarusottoRMP2013}. These polariton traps can be originated in planar microcavities due to intrinsic disorder~\cite{Baas2008,Lagoudakis2010}, can be fabricated by deep etching of microcavity pillar-like structures~\cite{Abbarchi2013,Galbiati2012}, created by micro-structuring the spacer of a microcavity in-between growth steps performed by molecular beam epitaxy (MBE)~\cite{Winkler2015,Kuznetsov2018,Chafatinos2020}, or using laser-induced effective potentials~\cite{Ohadi2018}. The same kind of semiconductor layered structures can be designed to perform as an acoustic phonon cavity with confined vibrational states~\cite{Trigo2002}. Similar to polaritons, these confined phonon states can be engineered as molecular-like levels in double structures or to form tailored engineered phonon bands in lattices~\cite{Kimura2007}. Interestingly, the same sequence of layer thicknesses designed to optimize a photon cavity leads to an optimum phonon cavity, with perfect overlap of the photon and phonon fields allowing for the conception of high-performance optomechanical resonators operating in the tens of GHz range~\cite{Fainstein2013,Anguiano2017,Lamberti2017,Zambon2022,Sesin2023}. This allows for the mechanically induced modulation of the inter-site coupling in polaritonic lattices.

Two mechanisms contribute to the photon-phonon interaction in these DBR-based microcavity structures, namely, radiation pressure (related to the phonon-induced displacement of the cavity interfaces) and electrostriction (related to the phonon-induced shift of electronic levels induced by deformation potential interaction~\cite{Baker2014}). These two mechanisms are of similar magnitude far away from electronic resonances, but electrostriction can be enhanced by several orders of magnitude by excitonic resonances which are intrinsic to the exciton-photon strong-coupling regime ~\cite{Jusserand2015,Scherbakov2022,Zambon2022,Sesin2023}.

Indeed, polaritons in the proposed resonators originate from the strong resonant coupling between the confined photon modes and excitons in quantum wells that are embedded in the cavity spacer~\cite{CarusottoRMP2013}. Thus, this system naturally becomes a playing ground to also exploit exciton resonances to enhance the optomechanical interaction. Typically, to maximize the photon-exciton strong coupling, the quantum wells (QWs) are positioned at anti-nodes of the cavity-confined electromagnetic field ($E$). In order to display strong optomechanical phenomena as considered in this work, however, the embedded QWs need to be displaced away from the position of the maximum cavity optical field. It turns out that at the position of the maximum of the optical field the strain associated with the confined phonon field ($s$) is zero, and thus the exciton-mediated polariton-phonon coupling vanishes.  Exciton-photon strong-coupling and optomechanical interactions are both optimized simultaneously if the QWs are positioned at the spacer positions where the product $s|E|^2$ is maximum. The parameters we will use in our model calculations correspond to realistic values for these optimized optomechanical cavity-polariton resonators~\cite{Chafatinos2020,Reynoso2022,Kuznetsov2021,Zambon2022,Chafatinos2022}.

Our model consists, as schematized in Fig.~\ref{Fig1}, of two coupled polariton modes ($j=1,2$)  and the corresponding reservoir's densities as~\cite{Wouters2008} 
\begin{eqnarray}
\nonumber
i \dot{\psi}_j&=&(\omega_j+U_j^{\textrm{P}} |\psi_j|^2+U_j^{\textrm{R}} n_j)\psi_j-J\psi_{3-j}\nonumber\\ && +\frac{i}{2}(R\, n_j-\gamma)\psi_j\,, \nonumber\\
\dot{n}_j&=& P_j-\gamma_{\textrm{R}}\,n_j-R|\psi_j|^2 n_j\,.
\label{eq:GP}
\end{eqnarray}
Here $\omega_j$ is the bare frequency of the $j$ mode, $U^{\textrm{P}}_j$ and $U_j^{\textrm{R}}$ are the polariton-polariton and reservoir-polariton interaction coupling rates, respectively, $J$ describes a direct hopping rate term between modes [determined by the potential $V(x,t)$ in Fig.~\ref{Fig1}], $\gamma$ the polariton decay rate, and $R$ the stimulated loading from the reservoir. The dynamic of the latter is controlled by the pump power $P_j$, the excitonic decay rate $\gamma_{\textrm{R}}$, and the stimulated decay rate to the condensate. We notice here that since we are considering two separate modes (on different traps) the direct overlap between them can be assumed small and, consequently, the equation for the local value of the reservoir density only depends on the amplitude of the corresponding polariton mode. A more involved model, including a cross term (as would be the case for a single trap), could be easily included. In that case, synchronization due to the competition of the reservoir-mediated population of the modes is also possible \cite{Eastham2008}.

\section{A dynamical systems approach to the synchronization phase diagram}
\label{sec:J0}
\begin{figure}[t]
    \centering \includegraphics{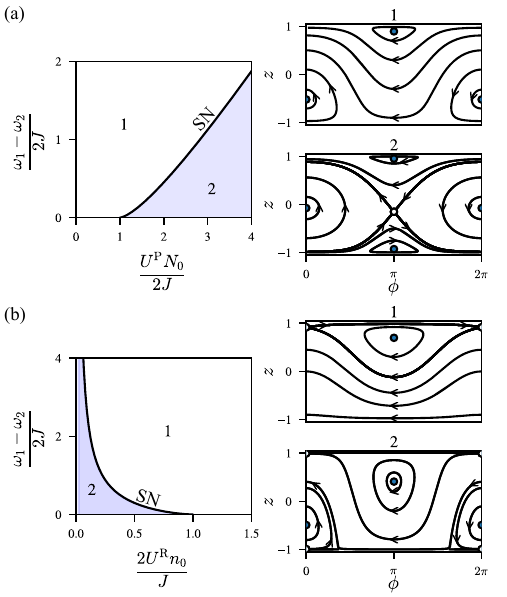}
    \caption{Bifurcation diagram for the conservative system. We show the phase portraits in the cylinder, i.e., $(\phi$ mod $2\pi, z)$. Empty dots represent saddle fixed points, blue dots non linear centers. (a) Polariton-polariton interaction only. (b) Reservoir-polariton interaction only. SN stands for saddle node.
    }
\label{Fig2}
\end{figure} 
We first consider previous developments in this domain, extending them later to the case of a dynamical mechanically-mediated inter-trap coupling. In this section, given the absence of a time-modulation induced by a coherent population of phonons, the inter-trap coupling $J$ is time independent and the system is known to display synchronization behavior~\cite{Wouters2008,Eastham2008,Eastham2021,Baas2008,Ohadi2018}. Even for this constant $J$ situation the full model from Eq.~\eqref{eq:GP} requires the analysis of a six-dimensional dynamical system. However, the experiments are usually done in conditions that allow further simplifications of the model. For this purpose, we assume that the populations of the reservoirs evolve in a faster timescale than the polaritons' dynamics. This allows us to perform an adiabatic approximation of the reservoirs' dynamics (setting $\dot{n}_j = 0$) and simplify two of the six dimensions. Additionally, for a pumping rate well above condensation threshold, such that the filling rate of the polariton condensates is much faster than the exciton decay ($R \left| \psi_j \right|^2 \gg \gamma_{\text{R}}$), we then have $n_j \simeq \frac{P_j}{R\left| \psi_j \right|^2}$.

\begin{figure*}[ht]
    \centering \includegraphics[width=1\linewidth]{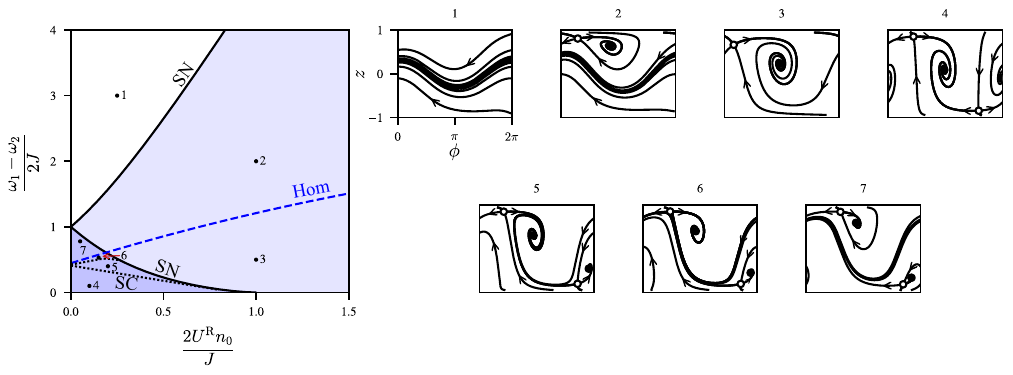}
    \caption{Bifurcation diagram of the model [(]Eq.~\eqref{eq: sistema_dinamico_zphi}] in the dissipative regime. The parameters are set to $\Gamma = 0.5$, $\alpha=0$.  The bifurcation curves define seven regions with qualitatively different dynamics. The insets show a representative phase portrait for each region. SN, Hom, and SC stand for saddle-node, homoclinic, and saddle-connection bifurcations, respectively (see text for details).
The vertical and horizontal axis in the right panels correspond to $z$ and $\phi$ mod $2\pi$, respectively. }
    \label{Fig3}
\end{figure*} 

Writing the mode's amplitudes as $\psi_j = \sqrt{N_j} e^{i \theta_j}$, we define two new variables $z(t) = \frac{N_1 (t) - N_2 (t)}{N_0}$, the relative population imbalance, and $\phi(t) = \theta_1 (t)  - \theta_2 (t)$, the condensates mode's phase difference, where $N_0$ is the total polariton population which the system approaches exponentially fast. For the symmetric pumping case $P_1 = P_2 = P$, the dynamical equations read as
\begin{equation}
    \begin{aligned}
    \frac{dz}{d\tau} &= \sqrt{1 - z^2} \sin{(\phi)} - \Gamma z\,, \\
    \frac{d\phi}{d\tau} &= -\frac{\omega_1 - \omega_2}{2J} - \alpha z
    + \sigma \frac{z}{1-z^2} - \frac{z}{\sqrt{1-z^2}} \cos{(\phi)}\,,
    \end{aligned}
    \label{eq: sistema_dinamico_zphi}
\end{equation}
where we have performed a scaling of the time $t=T \tau$ with $T=1/(2J)$ and defined the adimensional parameters $\Gamma \equiv \frac{\gamma}{2J}$, $\alpha \equiv \frac{U^{\mathrm{P}} N_0}{2J}$, and $\sigma \equiv \frac{2 U^{\mathrm{R}} n_{0}}{J}$, with $n_{0} \equiv \frac{P}{RN_{0}}$. The equations for $N_{1}+N_{2}$ and $\theta_{1}+\theta_{2}$ decouple from the equations for $z$ and $\phi$; this effectively reduces the dimensionality of the dynamical system from four to two.

We employ dynamical systems theory to identify qualitatively different solutions and synchronization conditions of this model, particularly as a function of the detuning between the polariton modes (difference between $\omega_2$ and $\omega_1$). In the synchronized state, the polariton modes settle into a configuration with nonzero population and a constant phase difference, corresponding to fixed points of the problem. These points $(\phi^{*}, z^{*})$ can be computed by solving the equation $f_{1, \pm}(z^*)=f_{2}(z^*)$ with 
\begin{eqnarray}    
    f_{1, \pm}(z) &\equiv& \frac{z}{1-z^{2}}\left(-\sigma \pm \sqrt{1-(1+\Gamma^{2})z^{2}}\right)\,,
    \nonumber \\ 
   f_{2}(z) &=& -\frac{\omega_1 - \omega_2}{2J} - \alpha z \,, 
    \label{eq: fixed_points}
\end{eqnarray}
under the condition $\Gamma \left |z^*\right | \leq {\sqrt{1-z^{*2}}}$. Once $z^{*}$ is computed, the phase can be obtained from the condition $\sin(\phi^{*}) = \frac{\Gamma z^{*}}{\sqrt{1-z^{*2}}}$. One can find curves in the parameter space that se\-pa\-ra\-te open regions where the necessary conditions for the existence of the fixed points can be satisfied. These are called bifurcation curves. Fixed points are born through a saddle-node (SN) bifurcation when $f_{1, \pm}(z)$ is tangent to $f_{2}(z)$ at $z=z^*$, that is $f_{1,\pm}'(z^*)=f_{2}'(z^*)$ with
\begin{eqnarray}
    \nonumber
        f_{1,\pm}'(z)&=& \frac{-\sigma (1+z^{2})\sqrt{1-(1+\Gamma^{2})z^{2}} \pm [1-z^{2}(1+2\Gamma^{2})]}{(1-z^{2})^{2} \sqrt{1-(1+\Gamma^{2})z^{2}} }\,, \\ f_{2}'(z)&=& - \alpha \,. 
    \label{eq: fixed_points_tangency}
 \end{eqnarray}
We separate our analysis in two scenarios, with and without dissipation.

\textit{Analysis of the model with no dissipation.} When $\Gamma = 0$, the system exhibits conservative dynamics, akin to quantum coherent tunneling in trapped Bose-Einstein condensates~\cite{Smerzi1997, Raghavan1999}.  Crucially, conservative systems lack \emph{attracting} fixed points~\cite{Strogatz1994}.  This translates to the inability of the polariton condensates to achieve a stable synchronized state in the absence of dissipation. In Fig.~\ref{Fig2} we show the bifurcation diagram and the phase portraits considering either the polariton-polariton interaction [$\alpha \neq 0$, $\sigma = 0$, Fig.~\ref{Fig2}(a)], or the reservoir-polariton interaction [$\alpha=0$, $\sigma \neq 0$, Fig. \ref{Fig2}(b)]. In both cases there is a region (denoted as region 1) with two fixed points. For the case with only polariton-polariton interaction (Fig.~\ref{Fig2}(a)) both fixed points are centers while in the reservoir-polariton case (Fig.~\ref{Fig2}(b)) there is a saddle and a center. Two new fixed points are born through a SN bifurcation (in region 2). In the first case the fixed points are born in $\phi=\pi$, while in the second one they appear in $\phi=0$. 

\textit{Analysis of the model with dissipation.} Figure \ref{Fig3} shows the bifurcation diagram for $\Gamma\neq0$~\cite{Lagoudakis2010,Abbarchi2013,Marino1999} and taking into account only the reservoir-polariton interaction---similar dynamics is found for a dominant polariton-polariton interaction. As the system is invariant under the transformation $(z, \phi, \frac{\omega_1-\omega_2}{2J}) \rightarrow (-z, -\phi, -\frac{\omega_1-\omega_2}{2J})$ we restrict our analysis to $\frac{\omega_1-\omega_2}{2J}>0$.

In region $1$, at high bare detuning, no fixed points exist and all initial conditions evolve to a stable limit cycle (the so-called Josephson oscillations). For the analytical conditions  described in Appendix~\ref{ap:SynchConditions} the system reaches region $2$ where two fixed points are born through a SN bifurcation. Interestingly, the presence of one stable fixed point implies that synchronization can appear in this region. If the system starts at the basin of attraction of the stable fixed point, the condensates synchronize, otherwise the system converges to a Josephson oscillation as in region $1$. This is consistent with Ref.~[\onlinecite{Wouters2008}]'s identification of a parameter region in which either synchronized or non-synchronized states could be reached depending on the initial conditions. Then, we find that at the dashed curve labeled as ``Hom", the limit cycle collides with the saddle point and disappears through a Homoclinic bifurcation. For bare detuning below this curve, at region $3$, the system synchronizes for all initial conditions, i.e., there exists a single stable fixed point. Indeed, these three regions dominate the map and are very similar to the main regions (or tongues) we discuss on the following sections (e.g., see Fig.~\ref{fig:Basins} below). 

For the sake of completeness, we also mention the existence of other regions in Fig.~\ref{Fig3}'s map which have a much smaller dominance in the parameter space. For example, in region $4$, two new fixed points are born in a SN bifurcation. In regions $4$, $5$ and $6$ the system undergoes two global bifurcations (Saddle-Connections labeled as "SC") in which the invariant manifolds of the saddle points are reorganized. In these regions we see that, even though the bare detuning favors a higher population in the first trap, the system can converge to a state in which the second condensate is more populated. In region $7$ we have two stable fixed points and a periodic orbit as attracting limit sets or attractors.

\section{Mechanically induced modulation in the coupling $J(t)$}
The motivation to theoretically study the effects of optomechanical dynamic modulation in the context of polariton condensate synchronization is based on a series of recent experimental works describing coupled phonon and polariton coherent phenomena in non-resonantly excited arrays of polariton micrometer-size traps. This begins with the observation of polariton-driven phonon lasing~\cite{Chafatinos2020}, i.e., the establishment of self-induced coherent mechanical oscillations of $\Omega\approx 20$GHz frequency, which were observed when a resonant condition was established tuning neighbor polariton condensates at energy differences corresponding to integer numbers of the phonon frequency $\Omega$. These phenomena were later explained due to optomechanical parametric oscillation induced by coupling between relatively distant traps~\cite{Reynoso2022} quadratically proportional to the phonon displacement operator, $\hat{x}=x_{zpf}(\hat{b}+\hat{b}^\dagger)$, as given by the following optomechanical interaction Hamiltonian,  
\begin{equation}
    H_{OM}=-\hbar g_n  (\hat{b}+\hat{b}^\dagger)^n  (\hat{\psi}_1^\dagger\hat{\psi}_2^{}+\hat{\psi}_2^\dagger \hat{\psi}_1^{})\,,
    \label{EQ:HintOM}
\end{equation}
taking $n=2$, where $g_2$ is the optomechanical second-order coupling, and $\hat{b}$ and $\hat{\psi}_i$ are the bosonic annihilation operators for the phonon and polariton modes, respectively. In principle, also phonon lasing involving an inter-trap optomechanical coupling that is linear in the phonon displacement, i.e., taking $n=1$ in Eq.~\eqref{EQ:HintOM}, can exist and dominate, for traps that are closer to each other so that there is a finite direct coupling, as previously reported for coupled photonic microdisks~\cite{Grudinin}. Very recently the experimental observation very stable locking of the relative frequencies between polariton traps in arrays was reported ~\cite{Chafatinos2022}. Very recently, the experimental observation of remarkably stable locking of the relative frequencies between polariton traps in arrays has been reported~\cite{Chafatinos2022}. This latter phenomenon resembles that of synchronization, but it contrasts with it in that the energies corresponding to neighbor polariton condensates do not become equal. Instead, they intriguingly get stabilized as the excitation power increases, with differences corresponding to integer multiples of the phonon frequency $\Omega$. All these previous experiments signal the self-induced establishment of coherent mechanical oscillation for continuous-wave and non-resonant laser excitation above the threshold for condensation in coupled polariton fluids. Notably, these resonant bulk acoustic waves in the GHz range can also be electrically generated~\cite{Kuznetsov2021,Crespo-Poveda2022}, opening the path to electrical high-frequency modulation of polariton condensate arrays.

To investigate these phenomena we now consider the case described by Eq.~\eqref{eq:GP} but when the coupling constant $J$ encodes the time-dependent modulation produced by a coherent mechanical wave of frequency $\Omega$ and period $T=2\pi/\Omega$. Namely,
\begin{eqnarray}
\nonumber
i \dot{\psi_1}&=&(\omega_1+U_1^{\textrm{P}} |\psi_1|^2+U_1^{\textrm{R}}\, n_1)\psi_1-J(t)\psi_{2} \nonumber \\ && +\frac{i}{2}(R\, n_1-\gamma)\psi_1\,, \nonumber\\
i \dot{\psi_2}&=&(\omega_2+U_2^{\textrm{P}} |\psi_2|^2+U_2^{\textrm{R}}\, n_2)\psi_2-\left(J(t)\right)^*\psi_{1}+\nonumber \\ && +\frac{i}{2}(R\, n_2-\gamma)\psi_2\,. 
\label{EQ:GP_ph}
\end{eqnarray}
In the case of relatively distant traps the direct coupling can be considered null between traps ground states (as e.g. described by the experiments in Ref.~\onlinecite{Chafatinos2022}). It then becomes relevant an inter-trap coupling that is quadratic in the phonon displacement (and which requires an excited polariton state that extends between the two traps~\cite{Reynoso2022}), leading to a time dependent hopping $J(t)=J_2 (e^{i 2 \Omega t}+e^{-i 2 \Omega t}+2)$ with $J_2=g_2 n_b$.  This follows from assuming the presence of a coherent population of $n_b$ phonons, i.e., $b(t)+b(t)^* =2\sqrt{n_b} \cos(\Omega t)$, and the interaction Hamiltonian of Eq.~\eqref{EQ:HintOM} with $n=2$. The periodicity in this case is $T/2$. On the other hand, if the two traps are closer to each other, direct inter-trap coupling can arise and Eq.~\eqref{EQ:HintOM} with $n=1$ (the linear case) would lead to the $T$-periodic $J(t)= J_1 (e^{i  \Omega t}+e^{-i  \Omega t})$ where $J_1=g_1 \sqrt{n_b}$.

\subsection{Rotating-wave approximation approach}
\label{sec:rwa}

The full implications of $J(t)$ affected by the phonon frequency $\Omega$, both the linear and quadratic cases, are presented in the next section. However, a first glimpse to its effects can be captured analytically by applying the rotating-wave approximation (RWA) to the equations of motion of Eq.~\eqref{EQ:GP_ph}. By approximating $J(t)\approx J_n e^{\pm i n \Omega t}$ we recover the time-independent equations for $J$ in Eq.~\eqref{eq:GP} of Sec.\ref{sec:J0} with $J_n \!\rightarrow\! J$ after writing 
\begin{equation}
    \psi_1=\sqrt{\rho_1}e^{-i\omega t+i\theta/2}~,~~\psi_2=\sqrt{\rho_2}e^{-i(\omega\pm n\Omega) t-i\theta/2}\,.
    \label{eq:ansatz}
\end{equation}
This RWA is valid for small values of $J_n$ and for detunings not too far from the resonance condition $\omega_2- \omega_1 = \pm n\Omega $ in which the non resonant terms of $J(t)$ can be neglected. Thus each Sec.~\ref{sec:J0}'s $\omega$-frequency synchronization case here translate into the two modes oscillating phase-locked with distinct frequencies. Explicitly, the system evolves in a mechanically induced {\em asynchronous} locking state, i.e., with the dressed frequencies of the two polariton modes appearing separated by $n\Omega$ or by $-n\Omega$. In addition, the zero-frequency contribution of the quadratic case, after neglecting the $\pm 2\Omega$ oscillating terms, directly fits Eq.~\eqref{eq:GP} leading to a synchronization region with $J=2J_2$.

Importantly, the locking regions behave exactly as the synchronization zone in Sec.~\ref{sec:J0}, i.e., their widths can be enhanced by the polariton-polariton or reservoir-polariton interactions and by $J_n/\gamma$. However, given the replacement $\omega_2- \omega_1 \rightarrow \omega_2- \omega_1 \pm  n\Omega$, the zone is located in a detuning shifted $\pm n\Omega$ with respect to the synchronization case. This RWA argument readily predicts that the linear case, since its spectral content is $\{-\Omega,\Omega\}$, produces two locking regions, or so-called tongues, shifted $\pm\Omega$ from zero detuning. Similarly, the quadratic case, with $\in\{-2\Omega,0,2\Omega\}$, generates a central zero-detuning synchronization tongue and two locking regions shifted in detuning $\pm2\Omega$.

In summary, the fixed points found in the case of Sec.~\ref{sec:J0} (time-independent $J$) justify, via the RWA and Eq.~\eqref{eq:ansatz}, locked states for the phonon-modulated $J(t)$ cases. In these states each polariton mode develops a single-frequency peak (without any sidebands) appearing separated by $\ell\Omega$, with $\ell$ defined as the locking number. For the linear case $\ell$ can be $\{-1,1\}$, while for the quadratic case it can be $\{-2,0,2\}$. Given a particular locking condition the associated $z(t)$ and $\phi(t)$ have the simple form
\begin{equation}
    z(t)=z^{*} ~,\phi(t)=\ell\Omega t +\phi^{*}\,.
    \label{EQ:rwaZPHI}
\end{equation}
As $\ell\Omega T=2\pi\ell$ the phase difference winds $\ell$ times after a single period of the drive, in this way, $\ell$ behaves as a winding ratio. When moving to the rotating frame in which the phase due to the locking rate is discounted the resulting trajectory, $(\phi(t)-\ell\Omega t,z(t))$, evolves to an attracting fixed point, $(\phi^*,z^*)$, exactly as in Sec.~\ref{sec:J0}. In the following sections, by using the full model, we show that for large interactions and/or large $J_n$ (with respect to $\Omega$) the system evolution deviates from this simple RWA dynamics. 


Interestingly, asynchronous locking behavior of the type presented here has recently been 
reported in a completely different setting. The Great Kiskadee ({\it Pitangus sulphuratus}), a bird from the Americas, remarkably achieves frequency locking between its two vocal chords.  The mechanics underlying this bird's song production can be described by non-linear dynamical equations remarkably similar to those used above for the coupled condensates~\cite{Doppler2020}.

\subsection{Full model and numerical simulations}
\label{sec:fullmodel}

\begin{figure*}[!!!ttt]
    \begin{center}
    \includegraphics[trim = 0mm 0mm 0mm 0mm, clip=true, keepaspectratio=true, width=1.9\columnwidth, angle=0]{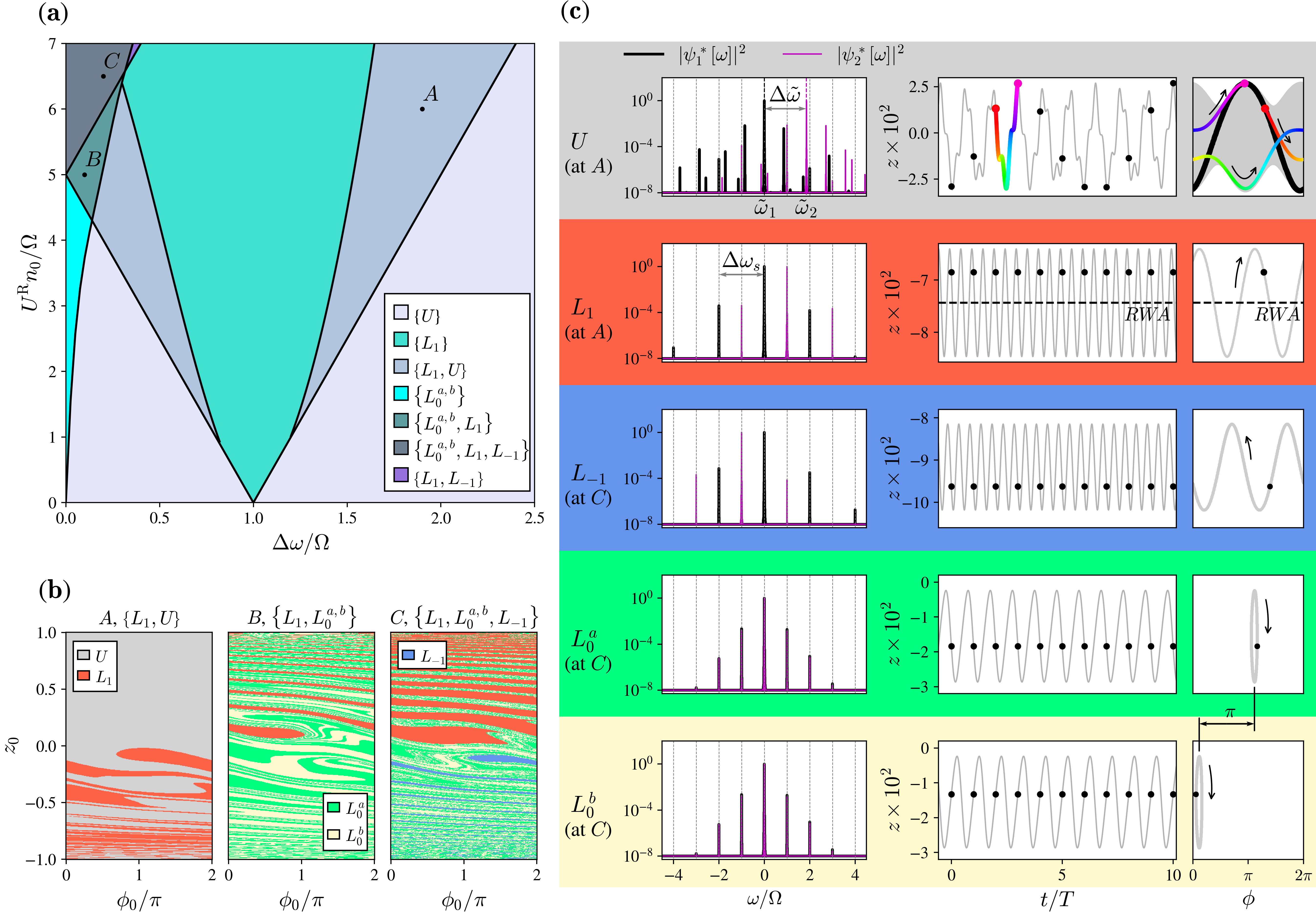}
    \end{center}
    \vspace{-0.5 cm}
\caption{(a) Regions having different asynchronous locking attractors, listed in brackets (see legend), for the linear case, in the parameter space of bare detuning, $\Delta\omega$, and reservoir-polariton interaction, $U^\mathrm{R} n_0$. Here $\gamma=0.2\Omega$, $J_1=\Omega/100$, $U^\mathrm{P}\bar{\rho}=0$ while the non-resonant driving is assumed balanced and in the strongly condensed limit, i.e., $P_1=P_2\gg P_{th}$. (b) Color-coded basins of attraction obtained by sweeping initial conditions $(\phi_0,z_0)$ for the three exemplary points marked in (a). The complexity of the basins appears because points $\mathrm{A}$, $\mathrm{B}$ and $\mathrm{C}$ correspond to 2-, 3-, and 4-attractor regions, respectively. For each of the five types of steady states, a row in (c) shows the following: left - the steady-state power spectra of the two modes, $|\psi_j[\omega]|^2$, with global frequency shifts removed for grid alignment (the dressed detuning $\Delta\widetilde{\omega}$ is obtained from $\tilde{\omega}_j$, the frequency difference between mode peak positions); center - the time dependence of $z(t)$ (gray line) with the Poincar\'e map period sampling (black dots); right - the $(\phi,z)$ phase-space Poincar\'e section (black) and full attractor (gray) with time-evolution arrows. Note that in all panels labels $L_\ell$, with $\ell=\{1,-1,0\}$, denote solutions having constant dressed detuning $\Delta\widetilde{\omega}=\ell\Omega$ which remain \emph{locked} in large regions of the parameter space. In contrast, for \emph{unlocked} solutions, labeled $U$, the dressed detuning varies smoothly (see Fig.~\ref{fig:Linear}) with the parameters. Labels $L^a_0$ and $L^b_0$ denote two nonequivalent attractors for the locking at $\ell=0$ differing, as shown in (c), by a $\pi$-shifted trajectory in $(\phi,z)$. Within the RWA, i.e., for $J(t)=J_1e^{i\Omega t}$, only the $\{L_1\}$ and the two $\{L_1,U\}$ locking regions are present; in (c) such $L_1$ solution evolves along the dashed line, while for each mode only the main peak of the spectrum would remain [see Eq.~\eqref{EQ:rwaZPHI}].}
\label{fig:Basins}
\end{figure*}

In this section we explore the polaritons' dynamics beyond the RWA. To do this we numerically solve the dynamics described by Eq.~\eqref{EQ:GP_ph} and analyze its steady states. Given that the state-of-the-art measurement techniques allow for sub-GHz resolution spectra \cite{Rozas2014}, we focus on the spectral properties of these signals. Importantly, such measurements have enough spatial resolution to identify the signals coming from each trap; this means experimental access to each $\psi^*_j[\omega]$, the Fourier transforms of each $\psi^*_j(t)$. To quantitatively and qualitatively classify the different situations we use the \emph{dressed} frequency detuning between the two modes as $\Delta\tilde\omega=\tilde\omega_2-\tilde\omega_1$, where each $\tilde{\omega}_j$ is defined as the frequency with maximum spectral weight in $\psi^*_j[\omega]$. As discussed above, for steady states reaching coherent phonon-induced locking conditions, this dressed detuning departs from changing smoothly with the bare detuning $\Delta\omega$ and, instead, within significant regions of the parameter space, it becomes linked to the phonon frequency as
\begin{equation}
 \Delta\tilde\omega=\ell\Omega\,.   
\end{equation}
In what follows, we use labels $L_\ell$ to denote such $\ell\Omega$ asynchronous locked states. For example, the synchronized state for $J(t)=J$ (without phonons) presented in Sec.~\ref{sec:J0} corresponds to a $\ell=0$ locking steady state, and so is labeled as $L_0$. As will be shown below, $\ell$ can also attain fractional values.

When solving Eq.~\eqref{EQ:GP_ph} we shall assume identical interaction strengths in both traps, i.e., $U_1^{\textrm{P}}=U_2^{\textrm{P}}=U^{\textrm{P}}$ and  $U_1^{\textrm{R}}=U_2^{\textrm{R}}=U^{\textrm{R}}$. As we see below this approximation suffices to explore the richness of the asynchronous locking phenomena. The non-resonant laser drives the exciton reservoir of the two traps providing the gain that populates the polariton modes. To quantify such gain, we define the polariton density that would be produced in the strongly condensed limit as $\bar{\rho}$, and the reservoir densities become $n_j= n_0/(P_{th}/P_j + |\psi_j|^2/\bar{\rho})$. With this notation the gain-loss factors in Eq.~\eqref{EQ:GP_ph} become
 \bea
 (R n_j-\gamma)&=&\gamma(n_j/n_0-1)
 \nonumber \\ &=&\gamma\left(\frac{1}{\frac{P_{th}}{P_j}+\frac{|\psi_j|^2}{\bar{\rho}}}-1\right)\,.
 \label{eq:gammaFull}
 \eea
 Note that, if the traps were decoupled, $J(t)=0$, gains and losses for each mode balance, the polariton populations reach $|\psi^\mathrm{eq}_j|^2=\bar{\rho}\xi_j^0/(\xi_j^0+1)$ where $\xi_j^0=P_j/P_{th}-1$ (as defined in Appendix \ref{ap:SynchConditions}) or $0$ below condensation. The effects of the reservoir-polariton and the polariton-polariton interactions are parameterized by the frequencies, $U^{\textrm{R}} n_0$ and  $U^{\textrm{P}} \bar{\rho}$, respectively. Unless otherwise stated, in what follows, $P_1=P_2=P$, and the condensates are driven in the strongly condensed limit, i.e., ${P_j}/P_{th}\gg1$. Thus $(R n_j-\gamma)\mapsto \gamma(\bar{\rho}/|\psi_j|^2 -1)$. In this limit in absence of $J(t)$ both modes would just condensate to  $|\psi^\mathrm{eq}_j|^2=\bar{\rho}$ and $z(t)=(|\psi_1|^2-|\psi_2|^2)/(|\psi_1|^2+|\psi_2|^2)$ would vanish in the steady state. Note that in this case the frequency shift due to the interactions, $ U^{\textrm{P}} \bar\rho +U^{\textrm{R}} n_0$, is identical for both modes implying that the bare detuning [\emph{bare} with respect to the phonon effect entering in $J(t)$] between both modes is just $\Delta\omega=\omega_2-\omega_1$. 

\subsubsection{Integer $\ell$ lockings: Linear case}

In this section we investigate the main locking states having integer $\ell$ for the linear case beyond the RWA, that is considering in full $J(t)=J_1 (e^{i  \Omega t}+e^{-i  \Omega t})$. We sweep initial conditions $(\phi_0,z_0)$ generating the $t_0$ mode amplitudes $\psi_1(t_0)=\sqrt{\bar{\rho}(1+z_0)/2}$ and $\psi_2(t_0)=\sqrt{\bar{\rho}(1-z_0)/2}\mathrm{e}^{-\mathrm{i}\phi_0}$. The initial time is taken $t_0=-5000 T$ ensuring that at $t=0$ the steady state is already reached.


We present the results in Fig.~\ref{fig:Basins}, working with values of $J_1$ small with respect to $\Omega$ in order to avoid strong coupling effects. Indeed, moderate values of $J_1/\Omega$ are consistent with previous experimental and theoretical studies. Figure~\ref{fig:Basins}(a) shows a map in the parameter space of $\Delta\omega$ and $U^{\textrm{R}} n_0$ classifying regions according to the \textit{characteristic set} of solutions, or attractors, that the system can reach by sweeping the initial conditions.

First, we identify a main locking zone that appears centered at bare detuning $\Delta\omega=\Omega$. Within this region all initial conditions lead to the same type of locking attractor having $\ell=1$ and, therefore, its characteristic set is just $\{L_1\}$.  Indeed, within such region, that broadens in detuning as the interaction strength increases, the steady state is consistent with the prediction made with the RWA transformation that linked the synchronized state of the time-independent case to the $\ell=1$ asynchronous locking. However, due to effect of the counter-rotating term in $J(t)$, the attractor of the full linear case differs slightly from the RWA one of Eq.~\eqref{EQ:rwaZPHI} as evidenced from the time-dependent trajectories and frequency spectrum of a typical $L_1$ attractor shown in Fig.~\ref{fig:Basins}(c). The relative population imbalance $z(t)$ oscillates around its mean value thus departing from being constant as the RWA case (gray line). Furthermore, the latter oscillation is linked to the mode spectra presenting weak $2\Omega n$-separated sidebands, which are absent in the RWA solution.

Figure~\ref{fig:Basins}(c) also includes, in black dots, the Poincar\'{e} map in $(\phi,z)$ obtained by period-sampling (with $T=2\pi/\Omega$) the steady-state solution as sketched in the $z(t)$ dependence. For the $L_1$ attractor this map generates a single-point Poincar\'{e} section which is a reflection of the $T$-periodicity of this particular attractor. The gray line shows the full trajectory of the attractor. By noticing the arrow direction one can see that $\phi$ advances $2\pi$ after a single period, which is consistent with $\ell=+1$. Such phase advance follows from the definition of asynchronous locking and can be discounted by moving to a rotating frame at the locking rate $\ell\Omega$, i.e.,   
\begin{equation}
\tilde{\phi}_\ell (t)\equiv  \phi(t) - \ell\Omega t \,. 
\label{eq:phi_ell}
\end{equation}

The latter rotation, that can be applied in general to any $\ell$-locked solution, discounts all the phase that produces phase winding. The resulting phase, $\tilde{\phi}_\ell (t)$, becomes strictly (not mod-$2\pi$) time-periodic in a multiple of the phonon period $T$. As we discuss below also for non-integer $\ell$ cases, such period depends on the locking condition and on the linear or quadratic nature of the coherent phonon-induced modulation of the inter-trap coupling. Importantly, the resulting $(\tilde{\phi}_\ell(t),z(t))$ trajectories encode information of beyond-RWA phenomena as they depart from remaining in a fixed point [see Eq.~\eqref{EQ:rwaZPHI} after discounting $\ell\Omega t$].

\begin{figure*}[!!!ttt]
    \begin{center}
    \includegraphics[trim = 0mm 0mm 0mm 0mm, clip=true, keepaspectratio=true, width=1.7\columnwidth, angle=0]{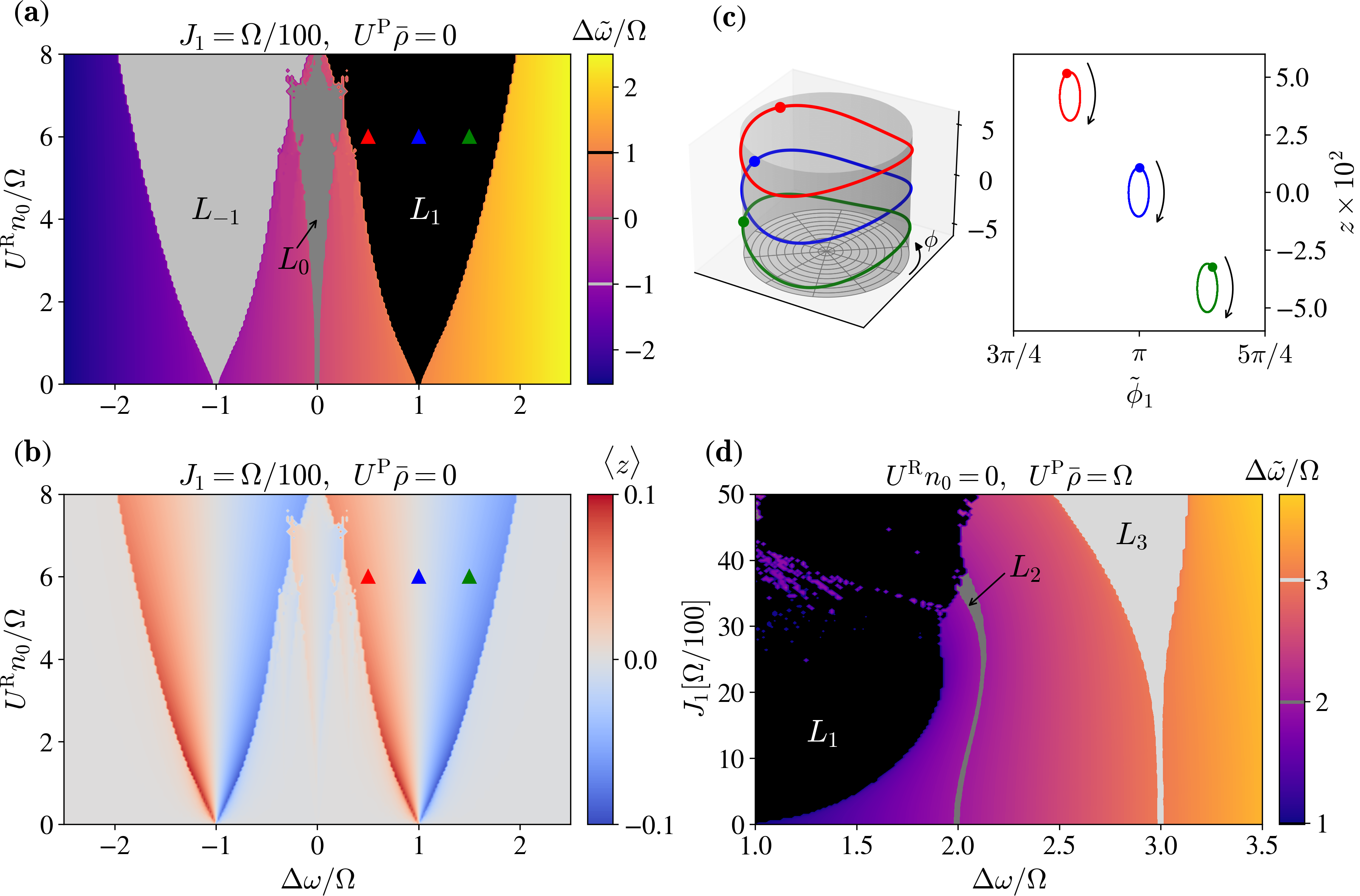}
    \end{center}
    \vspace{-0.5 cm}
\caption{Main integer locking regions for the linear case of $J(t)=J_1(\mathrm{e}^{\mathrm{i}\Omega t}+\mathrm{e}^{-\mathrm{i}\Omega t})$. Unless otherwise stated the driving and parameters are as given in Fig.~\ref{fig:Basins}. (a) Color map of the obtained dressed detuning $\Delta\tilde\omega$ (with main integer locked regions highlighted in gray scale) as a function of the bare detuning $\Delta\omega$ and of the reservoir-polariton interaction $U^\mathrm{R}n_0$ for $U^\mathrm{P}=0$ and $J_1/\Omega=0.01$ ; the corresponding map of the time-average of $z(t)$ color map is presented in (b). (c) Shows trajectories $z(\phi)$, for the three conditions inside the $\ell=1$ locking tongue marked in (a) and (b). These trajectories are represented both in 3D form using cylindrical coordinates for $\phi$, and in the 2D space in the rotating frame $(\tilde{\phi}_{\ell},z)$, i.e., after discounting the phase difference $\ell\Omega t$ that arises due to the $\ell\Omega$-frequency locking. In the latter, the remaining structure reveals the sideband content of the signals [as shown in the  spectrum for case $L_1$ in Fig.~\ref{fig:Basins}(c)]. For $U^\mathrm{P}\bar{\rho}=\Omega$ and $U^\mathrm{R}=0$ the color map in (d) shows $\Delta\tilde\omega$ as a function of $\Delta\omega$ and $J_1$ including the regime in which $J_1$ becomes a substantial fraction of $\Omega$ and integer $\ell>1$ locking conditions develop. As discussed in the text, odd $\ell$ are favored.}
\label{fig:Linear}
\end{figure*}

A second type of solution found in Fig.~\ref{fig:Basins}(a) is the \emph{unlocked} one, labeled $U$, for which the black points of the Poincar\'{e} map (also shown in Fig.~\ref{fig:Basins}(c)) fall on a sinusoidal-like shape evidencing quasi-periodicity as the map samples additional frequency scales different from the phonon frequency $\Omega$ or its harmonics. Indeed these scales arise (exactly within the RWA or approximately for the general case) from the spectral content of the stable limit cycle of the Josephson-type oscillations described in Sec.~\ref{sec:J0}. The dressed detuning of such unlocked solutions varies smoothly with parameters in contrast to the fixed $\Delta\tilde\omega$ in locked regions  (see the corresponding $\Delta\tilde\omega$ map in Fig.~\ref{fig:Linear}(a)). The gray lines connecting different points of such map (i.e., the full trajectory in Fig.~\ref{fig:Basins}(c)) cover a large region of the $(\phi,z)$ phase space (we color one $T$-lasting evolution between two points of the Poincar\'{e} map) in contrast to the simple trajectories found for locked attractors. This difference is also apparent from the much simpler spectra of the locked solutions for which the phonon provides a reference clock to the polaritons and not only the dressed detuning but also the sidebands spacing of each mode, $\Delta\omega_{s}$, become commensurable with $\Omega$.

The unlocked solutions appear alone in the region $\{U\}$, or in coexistence with the $L_1$ solution in the $\{L_1,U\}$ ones---see the basins of attraction in Fig.~\ref{fig:Basins}(b) for point A within a $\{L_1,U\}$ zone from which Fig.~\ref{fig:Basins}(c)  $L_1$ and $U$ attractors were taken. Indeed, with the RWA arguments of Sec.~\ref{sec:rwa} the regions $\{U\}$, $\{L_1,U\}$ , and $\{L_1\}$ can be connected to Fig.~\ref{Fig3}'s map zones 1, 2, and 3 of the time-independent $J$ case, respectively. In particular, the synchronization tongue for Fig.~\ref{Fig3}, centered at $\Delta\omega=0$, corresponds to Fig.~\ref{fig:Linear}(a)'s $L_1$ tongue since it appears $\Omega$-shifted in detuning by virtue of the time-dependent rotating transformation applied in Sec.~\ref{sec:rwa} to solve the RWA version of the coupling: $J(t)=J_1\mathrm{e}^{\mathrm{i}\Omega t}$. Similarly, the counter-rotating term of the linear coupling, $J_1\mathrm{e}^{-\mathrm{i}\Omega t}$, generates identical regions but involving $L_{-1}$, due to locking $\ell=-1$, centered at bare detuning $-\Omega$, see for example Fig.~\ref{fig:Linear}(a).      

The inclusion of the full $J(t)$ term generates a region in Fig.~\ref{fig:Basins}(a) with locking $\ell=0$ near zero detuning. In this zone we identify two distinct attractors, labeled $L_0^a$ and $L_0^b$, shown in Fig.~\ref{fig:Basins}(c). These attractors exhibit distinct characteristics compared to perfect synchronization (Sec.~\ref{sec:J0}): their phase space trajectories differ by $\pi$, and their spectra show $\Omega$-separated sidebands. Notably, increasing the interaction $U^{\textrm{R}} n_0$ strengthens this locking region. Crucially, this $\ell=0$ locking condition emerges between the two main locking regions predicted by RWA ($\ell=-1$ and $\ell=1$) in the bare detuning area. It corresponds to the first element in the Farey sequence of fractional $\ell$ that will be discussed later (Sec.~\ref{sec:frac}). For clarity, Fig.~\ref{fig:Basins}(a) omits a very thin intermediate region at the boundary between the $\{L_0^{a,b}\}$ zone and the unlocked region $\{U\}$ where both types of solutions can be reached.

Additionally, the map in Fig.~\ref{fig:Basins}(a) shows that as the interaction grows the $\{L_0^{a,b}\}$ region encounters the continuation of a $\{L_1,U\}$ one and what survives at their intersection does not contain the unlocked solution, i.e., the $\{L_0^{a,b},L_1\}$ region. The basins of attraction for point $B$ lying inside $\{L_0^{a,b},L_1\}$ are shown in Fig.~\ref{fig:Basins}(b). For interactions above $U^{\textrm{R}} n_0/\Omega\approx 5$, we see that the $\ell=-1$ attractor can be reached even at the positive detuning region and as it starts coexisting with other attractors in zones $\{L_1,L_{-1}\}$ and $\{L_0^{a,b},L_1,L_{-1}\}$. Figure~\ref{fig:Basins}(b) also includes the basins of attraction for point C in the latter zone, while the corresponding $L_0^a$, $L_0^b$ and $L_{-1}$ attractors' spectra and time-dependence are shown in three rows of Fig.~\ref{fig:Basins}(c).

It is worth mentioning that the existence of regions with different stable attractors naturally leads to hysteresis effects when a control parameter is experimentally changed slowly (with respect to $\gamma^{-1}$). Consider a parameter sweep starting from the $\{L_1\}$ region (locked solutions) and entering the $\{L_1,U\}$ region (coexistence). The system would likely remain in an $L_1$-like attractor until the sweep reaches the $\{U\}$ region, where the unlocked solution becomes the only stable state. Conversely, on the return path, the unlocked solution might persist until the sweep re-enters the $\{L_1\}$ region. Although in this work we do not discuss further hysteresis phenomena, given the multi-attractor landscape of our results, it is crucial to consider these effects for any particular parameter sweep applied in an experimental setting.

In what follows, as we focus on the regions that can be reached without tuning the initial condition, such as  $\{L_1\}$,  we can use a single initial condition. For this goal we explore the linear case by sweeping the most relevant parameters, including changing the balance of the driving conditions (see details in Appendix~\ref{ap:linear}). The main observations above for the integer-$\ell$ lockings remain. Namely, that the linear case is dominated by the $\ell=\pm 1$ lockings centered at bare detuning $\pm\ell\Omega$. And, as expected from Sec.~\ref{sec:J0} and the RWA arguments in Sec.~\ref{sec:rwa}, the widths of these tongues are enhanced by $J_1$, $\gamma^{-1}$, and more drastically, by the polariton-polariton and polariton-reservoir interaction strengths.

To conclude this section we investigate the dressed detuning and the mean value of the relative population imbalance within one of the main locking regions as well as the higher $|\ell|>1$ locking regions appearing in the strong $J_1/\Omega$ regime. Figure~\ref{fig:Linear}(a) presents the dressed detuning color map for the case introduced in Fig.~\ref{fig:Basins} having  $U^{\textrm{P}}\bar\rho=0$, $\gamma=0.2\Omega$ and $J_1=\Omega/100$.  The main locking regions show fixed steps where $\Delta\tilde\omega=\ell\Omega$ with $\ell$ equal to $1$ and $-1$ and the corresponding bare detuning widths grow with the interaction strength $U^{\textrm{R}}n_0$;  the same is true for the thinner locking tongue with $\ell=0$. Figure~\ref{fig:Linear}(b) shows the associated map for the time-average of $z(t)$, $\langle z\rangle$,  which becomes nonzero for bare detuning away from the center of each tongue. Such trend in $\langle z \rangle$ is also observed in the trajectories for the three exemplary points within the $L_1$ tongue shown in Fig.~\ref{fig:Linear}(c), in particular the one at the middle of the tongue has $\langle z \rangle=0$.

This can be understood as a requirement for the dressed detuning to be locked since, for $P_1=P_2\gg P_{th}$,   
\bea
\Delta\tilde\omega &\approx &\Delta\omega + U^{\textrm{P}} \langle|\psi_2|^2-|\psi_1|^2\rangle + U^{\textrm{R}} n_0 \langle \frac{\bar\rho}{|\psi_2|^2}-\frac{\bar\rho}{|\psi_1|^2}  \rangle \nonumber \\
&\approx & \Delta\omega - 2 U^{\textrm{P}}\bar\rho \langle z\rangle + 2 U^{\textrm{R}} n_0 \langle z\rangle. 
\label{eq:DressedDet}
\eea
Note that Eq.~\eqref{eq:DressedDet}, which considers only the contribution of the diagonal terms in the equations of motion, reasonably approximates the dressed detuning provided $J_1 \ll U^{\textrm{R}} n_0$ and/or $J_1 \ll U^{\textrm{P}} \bar{\rho}$. Since in this case only $U^{\textrm{R}} n_0$ is nonzero, it follows that $\langle z\rangle < 0$ ($>0$) for bare detunings $\Delta\omega - \ell\Omega > 0$ ($<0$), i.e., in the superior (inferior) half of the corresponding $\ell$-locking tongue. Equation~\eqref{eq:DressedDet} also reflects that to generate locking inside a given tongue, the two interactions require opposite sign profiles of $\langle z\rangle$ (see Appendix~\ref{ap:linear}). Thus, the enhancement due to interactions is absent when $U^{\textrm{P}}\bar{\rho} = U^{\textrm{R}} n_0$. However, this is not a problem for realizing the locking phenomena because this condition is generally not fulfilled.

A natural way to increase the width of the locking regions is by increasing $J_1$, which is achievable by growing the opto-mechanical coupling or the coherent population of the phonon mode. This is clear from the RWA mapping to the time-independent $J$ above and corroborated for the full model in Appendix~\ref{ap:linear}. This is, of course, a positive feature for the experimental observation of the effect even when the interaction is moderate. For very strong coupling and/or strong interactions chaotic behavior can appear as it is discussed below. At larger bare detunings, however, strong $J_1$ amplitudes can enable non-chaotic higher-order asynchronous locking conditions, e.g., see $\ell=3$ tongue in Fig~\ref{fig:Linear}(d). The fact that even-order $\ell$ are virtually absent can be understood by noting that $J(t)$ weights the component of the Hamiltonian proportional to a $\sigma_x$-like Pauli matrix operator in the subspace of modes $1$ and $2$. While odd orders lead to the mode mixing operator $\sigma_x$, the even orders of $\sigma_x$ are proportional to the identity operator. It follows that, since the identity operates equally on both modes, even orders of the $J(t)$-weighted operator do not produce nontrivial inter-mode behavior, including frequency locking.


\begin{figure}[!!!ttt]
    \begin{center}
    \includegraphics[trim = 0mm 0mm 0mm 0mm, clip=true, keepaspectratio=true, width=1\columnwidth, angle=0]{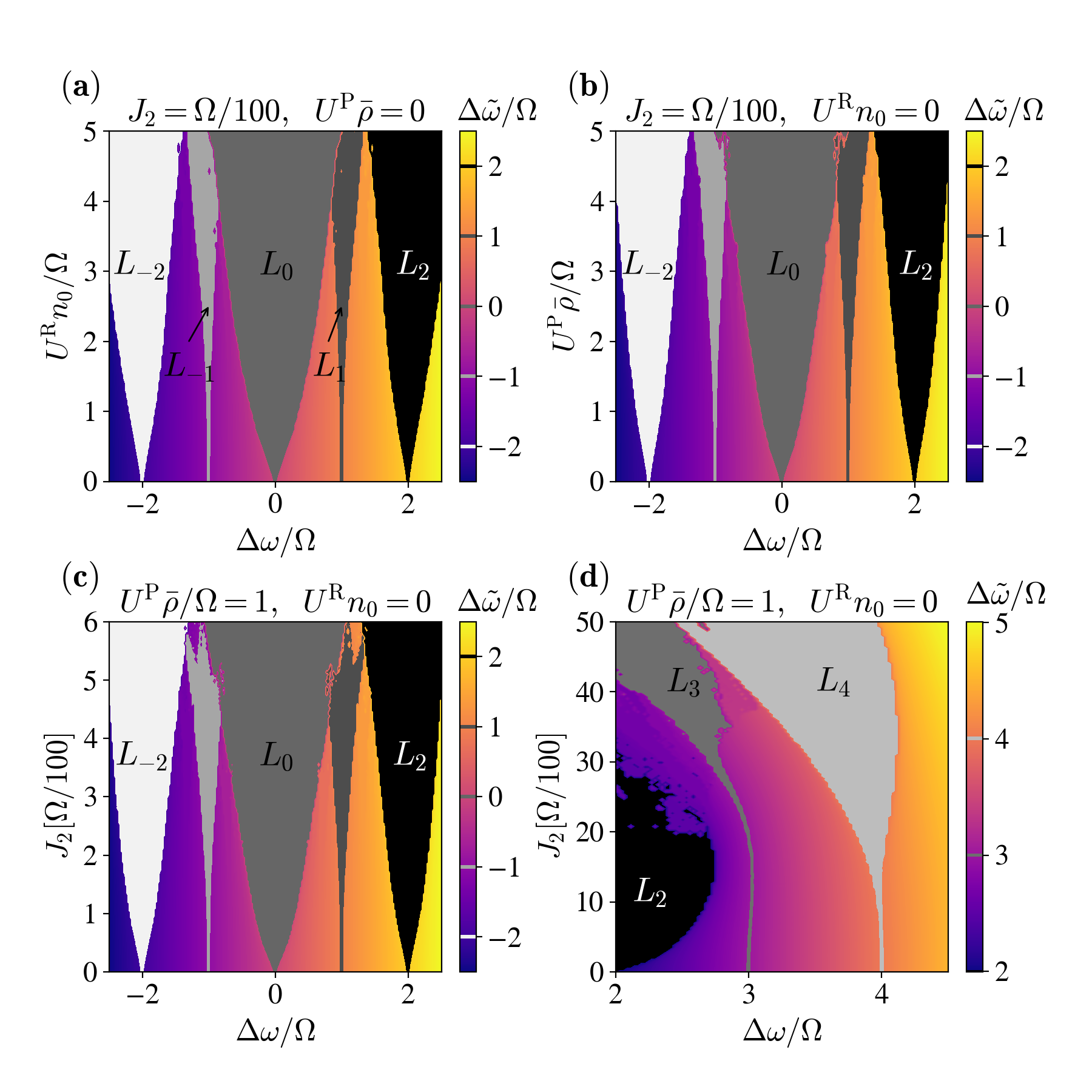}
    \end{center}
    \vspace{-0.5 cm}
\caption{
Behavior of the dominant integer locking regions for $J(t)$ arising due to a coherent phonon population in presence of a \emph{quadratic}-phonon intertrap optomechanical coupling for a single representative initial condition. Unless otherwise stated the parameters are as given in Fig.~\ref{fig:Basins}. Color maps of the obtained dressed detuning functionality (with main locked regions highlighted in gray scale) versus bare detuning $\Delta\omega$ and as a function of: (a) the reservoir-polariton interaction $U^\mathrm{R}n_0$ with $U^\mathrm{P}\bar{\rho}=0$; (b)  the polariton-polariton interaction $U^\mathrm{P}\bar{\rho}$ for $U^\mathrm{R}n_0=0$; and (c) [(d)] the amplitude of the coupling $J_2$ (for larger values of $J_2$ and $\Delta\omega$ showing higher-order locking) with $U^\mathrm{P}\bar{\rho}=\Omega$ and $U^\mathrm{R}n_0=0$. In contrast to the linear case of lockings $\pm1$ here the main integer tongues correspond to lockings $\ell\in\{2,0,-2\}$ while for larger amplitudes the $\ell=4$ locking dominates over $\ell=3$ of the linear case.
}
\label{fig:Quadratic}
\end{figure}

\subsubsection{Integer $\ell$ lockings: Quadratic case}
\label{sec:quad}

We now consider the quadratic case where $J(t) = J_2(e^{i 2\Omega t} + e^{-i 2\Omega t} + 2)$. The previously observed dependence of locking region width on interaction strength, coupling amplitude, and $\gamma^{-1}$ remains valid. However, due to the doubled frequency in the driving term, the main locking regions shift to $\ell = \pm 2$ and center around bare detuning values of $\pm 2\Omega$, as opposed to the $\ell = \pm 1$ regions observed in the linear case. This is evident in the $\Delta\tilde\omega$ maps of Fig.~\ref{fig:Quadratic}. Figures~\ref{fig:Quadratic}(a) and (b) show how these regions vary with interaction strength for a fixed value of $J_2/\Omega$, while Figs.~\ref{fig:Quadratic}(c) and (d) depict their dependence on $J_2/\Omega$ itself.

Another important difference arises from the mechanically-induced time-independent intermode-coupling of amplitude $2J_2$ contained in $J(t)$. This results in the emergence of a $\ell=0$ locking tongue, a direct manifestation of the synchronization due to time-independent hopping as described in Sec.~\ref{sec:J0}. This region, visible in Figs.~\ref{fig:Quadratic}(a)-\ref{fig:Quadratic}(c), appears centered at zero bare detuning and exhibits a wider width compared to those at $\pm2\Omega$. The reason for this wider width is that the $e^{\pm i 2\Omega t}$ factors in $J(t)$ have half the magnitude of the time-independent component.

\begin{figure*}[!!!ttt]
    \begin{center}
    \includegraphics[trim = 0mm 0mm 0mm 0mm, clip=true, keepaspectratio=true, width=1.9\columnwidth, angle=0]{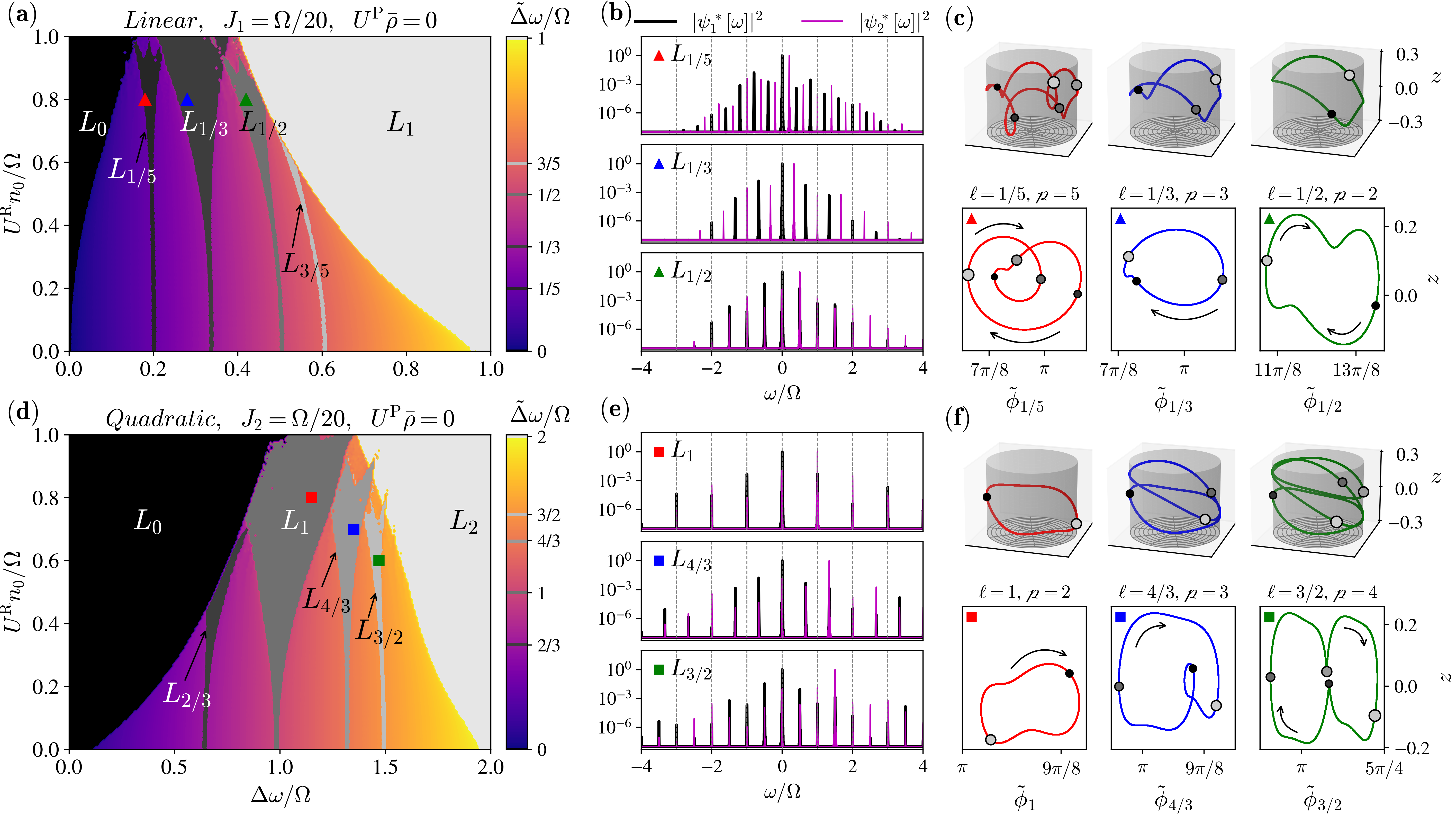}
    \end{center}
    \vspace{-0.5 cm}
\caption{
Exemplary fractional locking regions for both the linear and quadratic cases.  (a) [(d)] Color maps of the obtained dressed detuning functionality (with main locked regions highlighted in gray scale) versus bare detuning $\Delta\omega$ and as a function of the reservoir-polariton interaction $U^\mathrm{R}n_0$ with $U^\mathrm{P}\bar{\rho}=0$ for the linear case with $J_1=0.05\Omega$ [quadratic case with $J_2=0.05\Omega$]. Panels (b) and (c) show the spectra and trajectories, respectively, for three conditions corresponding to the fractional locking tongues $\ell\in\{\frac{1}{4},\frac{1}{3},\frac{1}{2}\}$ of the linear case in (a). The dots at these trajectories indicate the $T$-periodic sampling of the Poincar\'{e} section. Panels (e) and (f) show the spectra and trajectories, respectively, for three conditions corresponding to the fractional locking tongues $\ell\in\{2\times\frac{1}{2},2\times\frac{2}{3},2\times\frac{3}{4}\}$ of the quadratic case in (d).  As in the quadratic case the driving frequency is $2\Omega$ the dots at these trajectories indicate the $\frac{T}{2}$-periodic sampling of the Poincar\'{e} section. In both (c) and (f), the 2D trajectories are presented in the rotating frame corresponding to the locking $\ell$, i.e., $(\tilde\phi_\ell,z)$, while the full $(\phi,z)$ dependence is displayed in 3D-cylindrical coordinates.    
}
\label{fig:FractionsSpectra}
\end{figure*}

An additional feature of the quadratic case is the appearance of higher-amplitude nonlinear effects. These effects manifest as $\ell=\pm 1$ locking tongues with smaller widths, centered at bare detuning values of $\pm \Omega$. These zones, visible in Figs.~\ref{fig:Quadratic}(a)-\ref{fig:Quadratic}(c), arise due to the coexistence of counter-rotating terms in $J(t)$ and cannot be explained by the RWA arguments of Sec.~\ref{sec:rwa}. As discussed in Sec.~\ref{sec:frac} for fractional $\ell$ cases, these $\ell=\pm 1$ conditions share a conceptual similarity to the $\ell=0$ tongue observed in the linear case, which appears at a detuning midway between two frequency components of the driving term.

Furthermore, for larger $J_2/\Omega$,  Fig.~\ref{fig:Quadratic}(d) shows the emergence of a prominent $\ell=4$ locking tongue. This condition corresponds to twice the driving frequency ($2\Omega$) contrasting with the linear case, where only high-order tongues with odd multiples of the driving frequency were significant. Indeed, we have checked that in the quadratic case locking regions with $\ell=2n$ appear for both odd and even $n$. This can be understood by noting that in this case the odd powers of the interaction Hamiltonian---which can be thought as a Pauli matrix $\sigma_x^{2n+1}=\sigma_x$, i.e., non-trivially acting in the pseudo-spin space of trap $1$ and trap $2$---are weighted by a factor $[2J_2+2J_2\cos(2\Omega t)]^{2n+1}$ that includes even powers of the driving frequency $2\Omega$. Of course, the higher the order involved in each case, the larger the $J_2/\Omega$ required for obtaining a significant width of the corresponding $\ell=2n$ locking region. 

Figure~\ref{fig:Quadratic}(d) also reveals a less prominent $\ell=3$ locking region. The emergence of this condition, being at $3/2$ of the $2\Omega$ driving frequency of the quadratic case, indicates the necessity of a strong nonlinear effects. This requirement is evident in Fig.~\ref{fig:Quadratic}(d)'s $\ell=3$ locking region, for which the presence of reservoir-polariton interaction alone is insufficient, and a large coupling amplitude $J_2$, surpassing $30\%$ of $\Omega$, becomes necessary.

\subsubsection{Fractional $\ell$ lockings}
\label{sec:frac}

In this section, we show that the system develops lockings at fractions of the driving frequency. This can be seen, for example, as the phonon-modulated coupling amplitude $J_n$ is increased. Detailed color maps of the dressed detuning together with exemplary spectra and trajectories are presented in Fig.~\ref{fig:FractionsSpectra}. Figure~\ref{fig:FractionsSpectra}(a) focuses on the linear coupling case for bare detunings between $0$ and $\Omega$ with $J_1/\Omega=0.05$ and different values of the $U^{\textrm{R}} n_0$ interaction (the dependence as a function of $U^{\textrm{P}}\bar\rho$ is qualitatively equivalent). Notice that a cut along $\Delta\omega$ at a fixed $U^{\textrm{R}} n_0$ visits sequentially zones of constant $\Delta\tilde\omega=\ell\Omega$ including fractional $\ell$ values, i.e., it follows a devil's-staircase-like structure \cite{devilPhysToday}.

This intricate pattern arises due to the emergence of locking regions, $L_\ell$, having fractional values of $\ell$ subdividing the frequency range delimited by the main integer zones (the ones expected by the RWA arguments of Sec.~\ref{sec:rwa}), here $\ell=-1$ and $1$. Notably, the $L_0$ tongue, located at the center and discussed earlier (as the first locking condition beyond the RWA approximation), is prominent. Since the results are symmetrical with respect to this zone Fig.~\ref{fig:FractionsSpectra} only shows the positive detuning region to improve visualizing the relevant locking zones. Figure~\ref{fig:FractionsSpectra}(a) reveals the next most significant fractional regions (also called tongues) appearing in order of decreasing width: $L_\frac{1}{3}$, $L_\frac{1}{2}$, $L_\frac{1}{5}$, and $L_\frac{3}{5}$. Tongues with denominators larger than $5$ also exist but are not visible due to the limitations of the chosen resolution.

This type of Arnold tongue structure is a common signature of nonlinear effects in periodically driven oscillators when the driving term contains multiple frequency components. Attracting periodic solutions, including locking solutions, must arrange its Fourier components with frequency spacings compatible with the harmonic balance imposed by the spectral content of the excitation. In our case, the driving term of the linear case contains spectral weight at $-\Omega$ and $+\Omega$. Consequently, as the parameter space is explored, the dressed detuning has the potential of exhibiting locking at fractional values that subdivide the latter $2\Omega$ range into $n>1$ parts. In parameter space, such as bare detuning, interactions or coupling strengths, the emerging fractional tongues present widths that decrease with $n$. They require higher-order effects that typically do not manifest below threshold values that grow with $n$ (see Sec.~\ref{sec:doub}). However, such large $n$ zones can gain substantial widths, and become relevant in experiments, when the interactions and couplings are increased.

For example, as seen in Fig.~\ref{fig:FractionsSpectra}(a), the above introduced $\ell=0$ zone is the widest tongue of this type and it corresponds to $n=2$. Similarly, for $n=3$, we expect tongues with dressed detunings $-\Omega+2 k \Omega/3$ (where $k=1,2$). Indeed, the case with $k=2$ corresponds to the $\ell=\frac{1}{3}$ tongue observed in Fig.~\ref{fig:FractionsSpectra}(a). Importantly, the value of $n$, irrespective of $k$, determines the sideband frequency spacing of each mode. For the linear case we denote this spacing as $\delta\omega^{(1)}_n\equiv 2\Omega/n$. For example, in the $\ell=0$ case corresponding to $n=2$ [as shown in the characteristic double-peak structure of the dressed potential in Fig.~\ref{fig:Basins}(c)], the sideband spacing is $\Omega$. Similarly, the spectrum of the $\ell=1/3$ case in Fig.~\ref{fig:FractionsSpectra}(b) (corresponding to $n=3$) exhibits a sideband spacing of $\delta\omega^{(1)}=2\Omega/3$. 

More generally, the linear case presents fractional tongues with  $\ell=\ell^{(1)}_{n,k}$ with the definition $\ell^{(1)}_{n,k}=-1+2k/n$ with $k=1,\cdots,n-1$. 
Essentially, (if we also include the extremes of the range, $\pm 1$) the sequence of the allowed locking values $\ell^{(1)}_{n,k}$  corresponding to all $n\leq m$ configures a Farey sequence of order $m$, $F_m$, spanned from $-1$ to $1$ instead of from $0$ to $1$ \cite{detChaosBook,FareyTreesNatComm}. With this notation we can identify the remaining regions highlighted in Fig.~\ref{fig:FractionsSpectra}(a): the $\ell=\frac{1}{2}$ one corresponds to $\ell^{(1)}_{4,3}$ whereas $\ell=\frac{1}{5}$ and $\ell=\frac{3}{5}$ correspond to $\ell^{(1)}_{5,3}$ and $\ell^{(1)}_{5,4}$, respectively.

It is clear that an identical locking condition, $\ell$, can be obtained with different subdivisions of the range, e.g., $\ell^{(1)}_{n,k}=\ell^{(1)}_{n\times m,k\times  m}=\ell$, with $m>1$ positive integer. This corresponds to finer locking tongues inside wider ones sharing the same locking number $\ell$. Thus, by exploring the nearby parameter space or the initial conditions, the system can reach solutions having $m$ times finer sideband spacing. This feature is further explored for the period-doubling processes reported in Sec.~\ref{sec:doub} below. Of course, the case most likely to be observed (especially for smaller $J_1/\Omega$) is the one with smaller $n$, i.e., largest sideband spacing $\delta\omega^{(1)}_n=2\Omega/n$, as this is also the one having the largest widening of the locking zone.

The exemplary points marked in Fig.~\ref{fig:FractionsSpectra}(a) have $(\ell,n)\!\!\in\!\!\{(1/5,5),(1/3,3),(1/2,4)\}$. Their spectra, which are shown in Fig.~\ref{fig:FractionsSpectra}(b), confirm that the intra-mode side spacing coincides with the corresponding $\delta\omega^{(1)}_n = 2\Omega/n$. Additionally, for odd values of $n$, the spectral content of each mode exhibits weights at frequencies shifted from each other by $\delta\omega^{(1)}_n/2 = \Omega/n$. This translates to non-overlapping spectral components. In contrast, for even values of $n$, this inter-mode shift is absent [see the $L_0$ spectra in Fig.~\ref{fig:Basins}(c) corresponding to $n=2$]. This behavior arises because Eq.~\eqref{EQ:GP_ph}, for the linear $J(t)$ case, connects inter-mode frequencies separated by $\Omega$. Consequently, the spectral positions of different modes overlap for even-$n$ cases as the associated frequency spacing $\delta\omega^{(1)}_n$ perfectly divides $\Omega$. Conversely, for odd $n$, the intra-mode spacings $\delta\omega^{(1)}_n$ divide $2\Omega$ (by definition), but not the inter-mode separation of $\Omega$. This explains the observed $\delta\omega_n^{(1)}/2$ frequency shift between the spectral positions of the two modes.

For the same exemplary points Fig.~\ref{fig:FractionsSpectra}(c) shows their corresponding time-periodic attractors, either in a three-dimensional (3D) cylindrical representation of the $(\phi,z)$ trajectory or in the rotating frame $(\tilde\phi_\ell,z)$, i.e., by discounting the locking rate $\ell\Omega t$ to $\phi(t)$ as given in Eq.~\eqref{eq:phi_ell}. In both cases we include ordered dots indicating subsequent points of the Poincar\'{e} map obtained by sampling the attractor using the period of the driving, which for the linear $J(t)$ is $T=2\pi/\Omega$, until it completes its own period.

We use the number $\mathcal{p}$ to define the period of each attractor, $T_\mathcal{p}=\mathcal{p}T$, and find, as it can be verified by counting the dots of the Poincar\'{e} maps, that $\mathcal{p}=n$ for odd $n$ whereas $\mathcal{p}=n/2$ for even $n$. This is also reflected in the fact that for odd $n$ the trajectory in the rotating frame $(\tilde\phi_\ell,z)$ is visited twice to complete the period of the attractor $T_\mathcal{p}$ while, on the other hand, only once for even $n$. This is also directly linked to the above discussed $\delta\omega^{(1)}/2$ frequency shift found, for $n$ odd, between the two modes' spectral contents. As the trajectory follows from multiplication in the time domain of the two modes, the associated convolution in the frequency domain produces halved sideband spacing, $\delta\omega^{(1)}/2=\Omega/n$, with respect to the even $n$ case, thus justifying that $\mathcal{p}=n$ instead of $n/2$.

We define the period of each attractor, $T_\mathcal{p}$, as $\mathcal{p}T$. By analyzing the Poincar\'{e} maps (counting the dots), we find that $\mathcal{p} = n$ for odd $n$ and $\mathcal{p} = n/2$ for even $n$. This aligns with the observation in the rotating frame $(\tilde\phi_\ell,z)$: for odd $n$, the trajectory visits the phase space twice to complete the attractor period, while for even $n$, it visits only once. This behavior is directly linked to the previously discussed $\delta\omega^{(1)}/2$ frequency shift observed for odd $n$ between the two modes' spectra. Since the trajectory results from multiplying the two modes in the time domain, the corresponding convolution in the frequency domain leads to halved sideband spacing, $\delta\omega^{(1)}/2 = \Omega/n$, compared to the even $n$ case. This justifies why $\mathcal{p} = n$ for odd cases instead of $n/2$.

\textit{Quadratic case.} By changing the $U^{\textrm{R}} n_0$ interaction energy, as done in the linear case above, in Fig.~\ref{fig:FractionsSpectra}(d) we show the main fractional locking regions for the case of quadratic $J(t)$. We fix $J_2/\Omega=0.05$ and sweep the bare detuning from $0$ to $2\Omega$, i.e., in between two main locking zones ($\ell=0$ and $\ell=2$) predicted by the RWA.  As in this case $J(t)$ has spectral content at frequencies $-2\Omega$, $0$, and $2\Omega$, nearby frequencies also differ in $2\Omega$.  Then, in the displayed bare detuning region, expected dressed detuning lockings are expected to appear at $\Delta\tilde\omega=\ell\Omega$ where $\ell=\ell^{(2)}_{n,k}$ with the definition $\ell^{(2)}_{n,k}=2k/n$, $n>1$ and $k=1,..,n-1$. The associated intra-mode spectral sideband spacings are $\delta\omega_n^{(2)}\equiv 2\Omega/n$, identical to the ones of linear case $\delta\omega_n^{(1)}$.

In particular Fig.~\ref{fig:FractionsSpectra}(d) highlights the zones, in order of dominance,  $L_1$ due to $\ell^{(2)}_{2,1}$, $L_{4/3}$ due to $\ell^{(2)}_{3,2}$, $L_{2/3}$ due to $\ell^{(2)}_{3,1}$, and $L_{3/2}$ due to $\ell^{(2)}_{4,3}$. Since the spectral content of the quadratic $J(t)$ connects the two modes in frequency steps of $0$ or $\pm2\Omega$ here for both odd and even $n$ the positions with nonzero spectral weight of the two modes coincide, as it is shown for the spectra of three cases in Fig.~\ref{fig:FractionsSpectra}(e). Consequently, as the sideband spacing of the trajectory (involving mode $1$ and mode $2$ convolution in the frequency domain) is $2\Omega/n$ both $n$ parities lead to periodicity of $nT/2$. Considering that for the quadratic case the duplication of the driving frequency produces a $T/2$ intrinsic periodicity we redefine $T_\mathcal{p}=\mathcal{p}T/2$ and, consequently, $\mathcal{p}=n$ as it can be checked by counting the points of the Poincar\'{e} map, sampled at $T/2$ intervals, corresponding to the trajectories shown in Fig.~\ref{fig:FractionsSpectra}(f). 

Finally, it is apparent from Fig.~\ref{fig:FractionsSpectra}(f) that, regardless of the parity of $n$, the trajectories in the rotating frame, $(\tilde\phi_\ell,z)$, are visited only once per period $T_\mathcal{p}$. Indeed, this is expected in absence of a $\delta\omega_n^{(2)}/2$ frequency shift between the modes spectral positions (as the shift found above for $n$ odd in the linear case),

\begin{figure}[!!!ttt]
    \begin{center}
    \includegraphics[trim = 0mm 0mm 0mm 0mm, clip=true, keepaspectratio=true, width=1.0\columnwidth, angle=0]{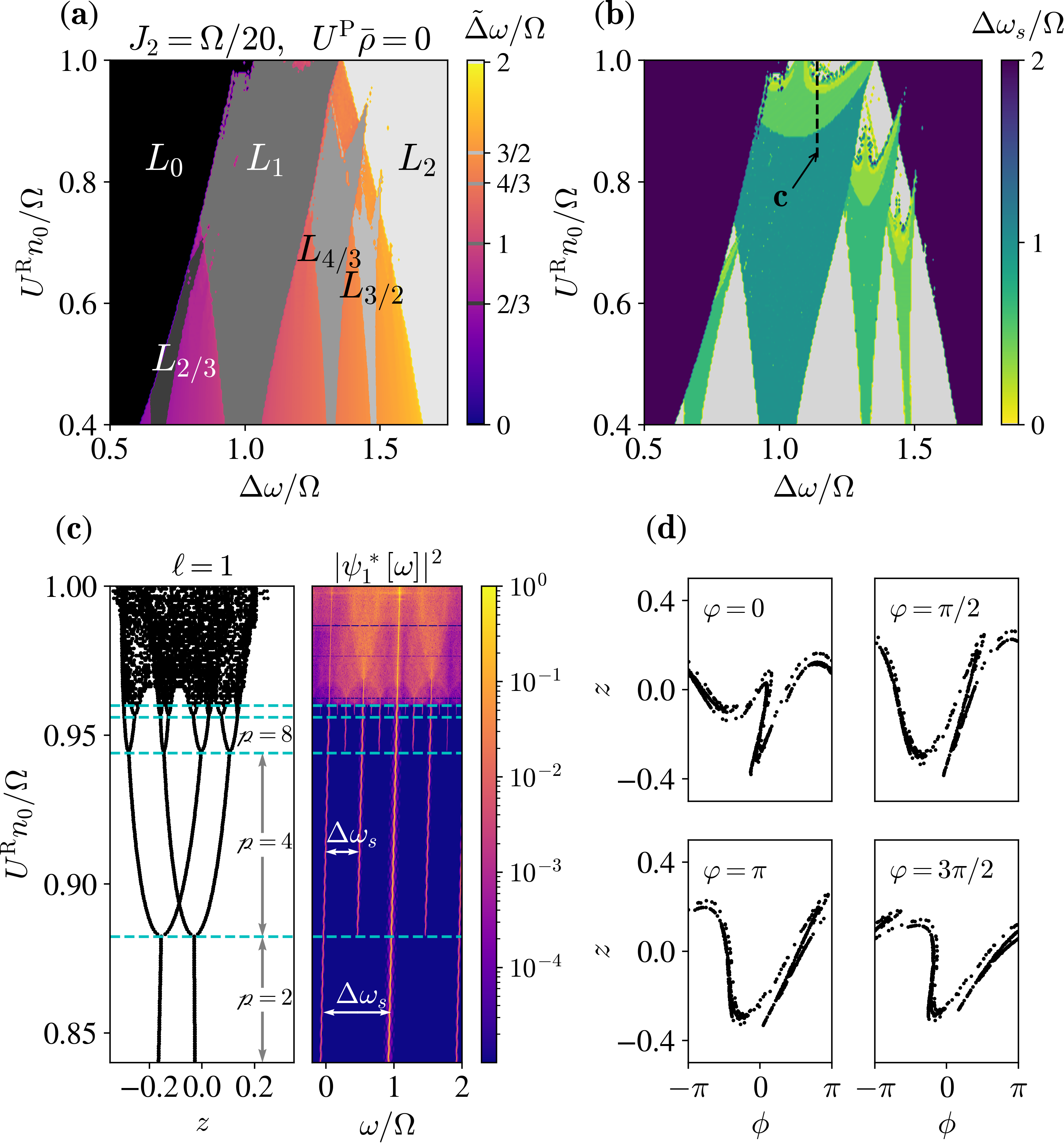}
    \end{center}
    \vspace{-0.5 cm}
\caption{
  (a)  Zoom of the dressed detuning color map given in Fig.~\ref{fig:FractionsSpectra}(c) for the quadratic case. (b) Corresponding color map of $\Delta\omega_s$: the sideband \emph{spacing} in the locking  steady states (the unlocked region is omitted in gray). For the main locking tongues, i.e., $\ell=0$ and $\ell=2$, the spacing $\Delta\omega_s=2\Omega$ is identical to driving of frequency $2\Omega$. Conversely, for fractional locking tongues the spacing $\Delta\omega_s$ undergoes frequency halvings as the interaction grows. Panel (c) shows the spectra for one of the modes along a vertical line inside the $\ell=1$ fractional locking tongue, evidencing that $\Delta\omega_s$ takes values $\frac{\Omega}{2^n}$ with $n>1$. Panel (d) values of $z(t)$ for the associated Poincar\'{e} section, sampled at $\frac{T}{2}$,  showing a sequence of period doubling bifurcations an the emergence of chaos. (d) Exemplary Poincar\'{e} sections of a chaotic attractor, $(\phi,z)$, taken at different intra-period phases (see text), illustrating the stretching and folding behavior.  }
\label{fig:FractionsChaos}
\end{figure}

\subsubsection{Period doubling and chaos}
\label{sec:doub}

As discussed above for the linear case, a given locking condition $\ell$ can arise with different spectral subdivisions. In the quadratic case, this relationship is captured by the expression $\ell^{(2)}_{n,k} = \ell^{(2)}_{n\times m, k\times m}$, where $m$ is any positive integer. These distinct $m$ attractors also exhibit unique characteristics in the time domain, as their period factor becomes $\mathcal{p} = n \times m$.

To delve deeper into this behavior, we examine not only the dressed detuning (Fig.~\ref{fig:FractionsChaos}(a)) but also the sideband spacing of a single mode in the steady state, denoted as $\Delta\omega_s$ (Fig.~\ref{fig:FractionsChaos}(b)). When the system is locked in a periodic attractor, this sideband spacing directly mirrors the intra-mode spectral spacing, expressed as $\delta\omega_{n\times m}^{(2)}=2\Omega/(n\times m)$. This allows us to readily determine the periodicity factor of the attractor using the relationship $\mathcal{p} = 2\Omega/\Delta\omega_s$.

Focusing on a specific fractional locking region and starting with a single initial condition, we observe that as the interaction parameter increases, $\Delta\omega_s$ undergoes a series of halving events, signaling period-doubling processes. This is clearly illustrated in Fig.~\ref{fig:FractionsChaos}(c), where we track the attractor properties along the line marked in Fig.~\ref{fig:FractionsChaos}(b) within the $\ell=1$ Arnold tongue. Consistency is observed between the halvings of $\Delta\omega_s$ (indicated by dashed horizontal lines) and the period doubling events ($\mathcal{p} = 2, 4, 8$) seen in the Poincar\'{e} map and spectra. These events correspond to the $\ell=1$ lockings $\ell^{(2)}_{2,1}$, $\ell^{(2)}_{4,2}$, and $\ell^{(2)}_{8,4}$, respectively, and exemplify a period-doubling route to chaos, characterized by increasing complexity in the Fourier transform.

Figure~\ref{fig:FractionsChaos}(d) showcases the Poincar\'{e} section for a chaotic state. Unlike periodic cases, chaotic Poincar\'{e} sections are no longer composed of a set of $\mathcal{p}$ points that are periodically revisited in phase space. By sampling the Poincar\'{e} map at times $t_n = (n + \varphi/(2\pi)) T/2$, we generate sections at different phases $\varphi$, which visualize the characteristic stretching-and-folding behavior often observed in chaotic attractors.

Despite the underlying chaos, as evidenced in the dressed detuning map [Fig.~\ref{fig:FractionsChaos}(a)], the main spectral components of the two modes may remain approximately locked to the $\ell$ value of the corresponding tongue. While such chaotic regimes hold potential for certain applications, they are generally undesirable for applications requiring coherent and predictable manipulation of polariton modes.

\section{Summary}

The dynamics of trapped polariton modes hosted in the discussed heterostructures can be dramatically affected by the presence of a coherent population of cavity phonons. Such population could arise self-consistently through the interaction with other degrees of freedom or from externally driven microwaves through surface or bulk acoustic wave generators. Indeed, experimental evidence of locking in the polariton dynamics due to self-induced phonon populations has been reported in Ref.~[\onlinecite{Chafatinos2022}], while Ref.~[\onlinecite{Kuznetsov2023}] also demonstrates polariton control using microwave-generated phonons. Depending on the situation the phonon displacement, with frequency $\Omega$, can linearly or quadratically modulate the induced inter-mode coupling $J(t)$.

To investigate the effect of this modulation on the polaritons we first revisited the situation in which the coupling is time-independent. We study the case in which two traps are non-resonantly driven by an external laser and the polariton modes condensate as independent self-sustained oscillators having their own frequencies. If the modes were uncoupled, the dressed detuning just follows the bare detuning $\Delta\omega$. A time-independent coupling $J$ between the traps allows for synchronizing the frequencies of the condensed modes in a region of bare detuning which is enlarged by either polariton-polariton or reservoir-polariton interactions as well as $J/\gamma$. 

Using a rotating-wave mapping we show that the latter synchronization effect, which is analytically formulated, can explain that, for coherent phonon-induced $J(t)$ modulation, there exist zones of bare detuning, centered at $\ell\Omega$, in which the main spectral lines of the two modes appear separated by a locked frequency of $\Delta\tilde\omega=\ell\Omega$, with $\ell=\pm 1$ or $\ell=0,\pm 2$ when the coupling contains the phonon linearly or quadratically, respectively. Therefore, within these bare detuning zones---which can be enlarged by the interactions, the $J(t)$ amplitude, and the polariton lifetime $\gamma^{-1}$---the spectral signature of the steady state is the locking of the dressed detuning, $\Delta\tilde\omega$, at values in multiples of the phonon frequencies.

Beyond the rotating-wave approximation several other effects develop and these were described in detail based on numerical simulations. First, the locking conditions persist despite the fact that the spectra of each mode present sidebands due to the presence of counter-rotating parts in $J(t)$. The initial conditions are important in regions of parameter space in which different locking attractors coexist, however, the main locking regions ($\ell=\pm 1$  for the linear case and $\ell=0,\pm 2$ for the quadratic case) present a core with a single attractor. This fact would produce hysteresis in experiments in which parameters are swept slowly through regions with different number of stable solutions.

Regarding the nature of the locked frequency states, the phase difference of the two condensates $\phi(t)$ has $\ell \Omega t$ contribution that can be discounted moving to a rotating frame. What remains is a structure related to the sideband complexity of the attractor, e.g., within the RWA this would be just a constant $\phi_0$. In the case of the relative imbalance between mode $1$ and $2$ occupations, $z(t)$, its time-average is zero at bare detunings $\ell \Omega$ (the center of the locking tongues) while it becomes nonzero away from such condition, with its sign depending on the dominance of polariton-polariton or reservoir-polariton interaction. This can in principle be modified by tuning the power, as the competence between $U^{\textrm{P}}\bar\rho$ and $U^{\textrm{R}} n_0$ adds a control knob to tune the polariton $\ell$-locked states.

In the case of stronger driving at larger bare detunings additional locking conditions arise, due to higher order of the driving interaction operator, corresponding to larger integers values of $|\ell|$; in particular, for the linear case only odd $\ell$ cases [e.g., $\ell=3$ in Fig.~\ref{fig:Linear}(i)] develop regions with significant bare detuning width. In addition, a prominent non-linear effect, that cannot be accounted either with the rotating-wave argument or due to higher orders of the coupling operator, is the onset of locking values $\Delta\tilde\omega=\ell\Omega$ intermediate between the main locking conditions. Such fractional conditions arrange in Arnold tongues, subdividing the $2\Omega$ ranges of the bare detuning $\Delta\omega$ in between the main locking regions and configuring typical devil's-staircase-like structures commonly found in periodic-driven nonlinear systems unveiling a Farey sequence of the corresponding $\ell$ values. Depending on the position of the tongue the lowest period of the attractor becomes a particular multiple of the $J(t)$ period. The bare detuning widths of those regions are also enhanced by the interactions and the amplitude of $J(t)$. Furthermore, if one of such parameters is increased, as the tongue widens the attractor sequentially undergoes period-doubling processes eventually reaching chaotic behavior.

In summary, coherent phonon-induced $J(t)$ produces a rich variety of integer and fractional locking conditions at the frequencies of the cavity phonons which in these systems range from $7-150$GHz. These two-trap polariton states are ideal for being detected with state-of-the-art spectroscopic methods involving interferometrically measuring visible light at hundreds of THz with spectral resolutions lower than $1$GHz. Indeed, Ref.~[\onlinecite{Chafatinos2022}] presents experimental evidence of multi-trap coherent phonon-induced integer $\ell$ locking in the polariton dynamics arising due to self-induced mechanical vibrations, while Ref.~[\onlinecite{Kuznetsov2023}] reports similar effects for the polariton modes including the case of using the resource of microwave-generated phonons. We expect that further experimental efforts can provide evidence of the fractional $\ell$ lockings, hopefully unveiling the first fractions in the Farey sequence and a devil's-staircase-like pattern in the measured dressed detuning.

\begin{acknowledgments}
We acknowledge partial financial support from the ANPCyT-FONCyT (Argentina) under grants PICT-2015-1063, PICT-2018-03255, PICT 2018-1509 and PICT 2019-0371, SECTyP UNCuyo 06/C053-T1. AAR acknowledges support by PAIDI 2020 Project No. P20-00548 with FEDER funds. GU thanks J. Tempere and M. Wouters for discussions during a research stay at UAntwerpen, partly funded by the Fund for Scientific Research-Flanders and N. Goldman for his hospitality at ULB. We thank D.H. Zanette from Instituto Balseiro for helpful insights.
\end{acknowledgments}

\appendix
\section{Conditions for the existence of a synchronized state}
\label{ap:SynchConditions}

The bifurcation diagram in Fig.~\ref{Fig3} expresses the physical fact that it is the Josephson flow that establishes synchronization, with larger inter-trap couplings $J$  required for increasing bare state detuning $\Delta \omega=\omega_2-\omega_1$. It also shows that for the model of Eq.~\eqref{eq:GP} synchronization can exist even without explicit non-linearities given by either $U_j^{\textrm{P}}$ or by $U_j^{\textrm{R}}$ (the reservoir dynamics $\dot{n}_j$ provides in this case the non-linearity in the dissipative term for $\dot{\psi}_j$). Notwithstanding this affirmation, as displayed in Fig.~\ref{Fig3} polariton-polariton and reservoir-polariton interactions strongly favor the emergence of the synchronized state by providing the means for the system to compensate for the intrinsic detuning $\Delta \omega$. The limit between unsynchronized and synchronized regions 1 and 2 in Fig.~\ref{Fig3} can be analytically defined searching for the conditions of existence for a solution of the form $\psi_j=\sqrt{\rho_j}e^{-i\omega t\pm i\theta/2}$ in which the two states share the same frequency $\omega$ (here the $+$ sign corresponds to $j=1$ and the $-$ sign to $j=2$), and with $\dot{n}_j=0$~\cite{Wouters2008}. Introducing this ansatz in Eq.~\eqref{eq:GP} the following set of algebraic equations follows:

\begin{eqnarray}
\nonumber
\frac{1}{\alpha}\frac{\left(\xi _1^0-\xi_1\right)}{\xi _1+1}&=&-2  J_g \sin(\theta )\\
\nonumber
\alpha\frac{ \left(\xi _2^0-\xi _2\right)}{\xi_2+1}&=&2 J_g \sin (\theta )\\
\nonumber
\alpha  J \cos (\theta )+\omega&=&\frac{U_1^{\textrm{R}} n_0\left(\xi _1^0+1\right)}{\xi _1+1}+\xi _1 \rho_0 U_1^{\textrm{P}}\\
\nonumber
\frac{J\cos(\theta )}{\alpha }+\omega&=&\frac{U_2^{\textrm{R}} n_0 \left(\xi _2^0+1\right)}{\xi _2+1}+\Delta\omega +\xi _2 \rho _0 U_2^{\textrm{P}}\\
\label{constrains}
\end{eqnarray}

where $\alpha=\sqrt{\rho_2/\rho_1}$ and we have introduced some dimensionless parameters so that:  $J_g=J/\gamma$, $\xi_j=\rho_j/\rho_0$, $\rho_0=\gamma_{\textrm{R}}/R$, $P_j=(1+\xi_j^0)P_\mathrm{th}$, $P_\mathrm{th}=\gamma \gamma_{\textrm{R}}/R$, $n_0=\gamma/R$. We also made use of the stationary solution for the reservoir, $n_j=P_j/(\gamma_{\textrm{R}}+R\rho_j)=n_0 \left(\xi_j^0+1\right)/(\xi_j+1)$.

Equations \eqref{constrains} need to be solved for $\xi_j$, $\omega$, and $\theta$. Notice that the first two allow for the determination of $\xi_j(\theta)$, independently of the interactions (for instance, $J=0$ implies $\xi_j=\xi_j^0$). While it is possible to find a general solution for $\xi_j(\theta)$, it involves a quartic polynomial with generic roots and, in practice, it is in general difficult to determine the one with physical meaning ($\xi_j,\alpha>0$).

However, an analytical condition for $\theta$ can be found if one assumes that $R \rho_j\gg \gamma_{\textrm{R}}\rightarrow \xi_j\gg 1$, i.e., $P_j\gg P_\mathrm{th}$, deep into the condensed regime \cite{Wouters2008}.  In that case, $\xi_j+1\rightarrow \xi_j$ in the above equations (and for consistency the same applies to $\xi_j^0$). Hence, from the first two equations in Eq.~\eqref{constrains} one gets
\begin{widetext}
\begin{eqnarray}
\xi_1(\theta)&=&\Theta(-\theta) g(\theta)+\Theta(\theta) \frac{(\xi _{1}^0)^2}{[1+4J_g^2\sin^2(\theta) ]g(\theta)}~,~~~\\
\xi_2(\theta)&=&\frac{\left(\xi_1(\theta)-\xi_1^0\right) \xi_2^0}{4 J_g^2 \xi_1(\theta) \sin^2(\theta)+\xi_1(\theta)-\xi_1^0}\,,
\end{eqnarray}
with
\begin{equation}
g(\theta)=\frac{(\xi _{1}^0)^2}{\xi_1^0+2 J_g \sin ^2(\theta ) \left(J_g \left(\xi _1^0+\xi _2^0\right)+\sqrt{\xi _1^0 \xi _2^0 \csc ^2(\theta )+J_g^2
   \left(\xi _1^0+\xi _2^0\right){}^2}\right)}\,,
\end{equation}
and $\Theta(x)$ the step function. The final equation that determines $\theta$ and the synchronization condition is then
\begin{eqnarray}
\nonumber
\Delta\omega &=&f(\theta)\\
\nonumber
&=& J\cos(\theta)\left(\frac{1}{\alpha(\theta)}-\alpha(\theta)\right)
+\frac{U_1^{\textrm{R}} n_0\xi _1^0}{\xi_1(\theta)}+ U_1^{\textrm{P}}\rho_0 \xi_1(\theta)-\frac{U_2^{\textrm{R}} n_0\xi_2^0}{\xi_2(\theta)}-U_2^{\textrm{P}}\rho_0 \xi_2(\theta)\,.\\
\label{Theta}
\end{eqnarray}  
\end{widetext}
Once $\theta$ is determined, the locking frequency is given by
\begin{equation}
\omega=\frac{\bar{\varepsilon}_1+\bar{\varepsilon}_2}{2}-\frac{J\cos(\theta)}{2}\left(\alpha+\frac{1}{\alpha}\right)\,,
\end{equation}
and since
\begin{equation}
    \bar{\varepsilon}_2-\bar{\varepsilon}_1=J \cos(\theta)\, \left(\frac{1}{\alpha}-\alpha\right)\,,
\end{equation}
we get
\begin{equation}
\omega=\bar{\varepsilon}_1-\alpha\, J\cos(\theta)\,.     
\end{equation}
Here $\bar{\varepsilon}_j=\omega_j+U_j^{\textrm{P}} \rho_0 \xi_j+U_j^{\textrm{R}}n_0 \frac{\xi_j^0}{\xi _j}$ are the trap frequencies dressed by the interactions. Notice that we allowed here for different pump powers on each trap, $\xi_1^0\neq\xi_2^0$. The resulting synchronization condition is identical to the one presented above in Fig.~\ref{Fig3}, except that the bifurcation diagram is not longer symmetric around $\Delta\omega=0$ when $\xi_1^0\neq\xi_2^0$.

\renewcommand\thefigure{\thesection.\arabic{figure}} 
\section{Linear case: In-depth exploration of the main integer locking regions}
\label{ap:linear}
\setcounter{figure}{0}

\begin{figure*}[!ht]
    \begin{center}
    \includegraphics[trim = 0mm 0mm 0mm 0mm, clip=true, keepaspectratio=true, width=1.9\columnwidth, angle=0]{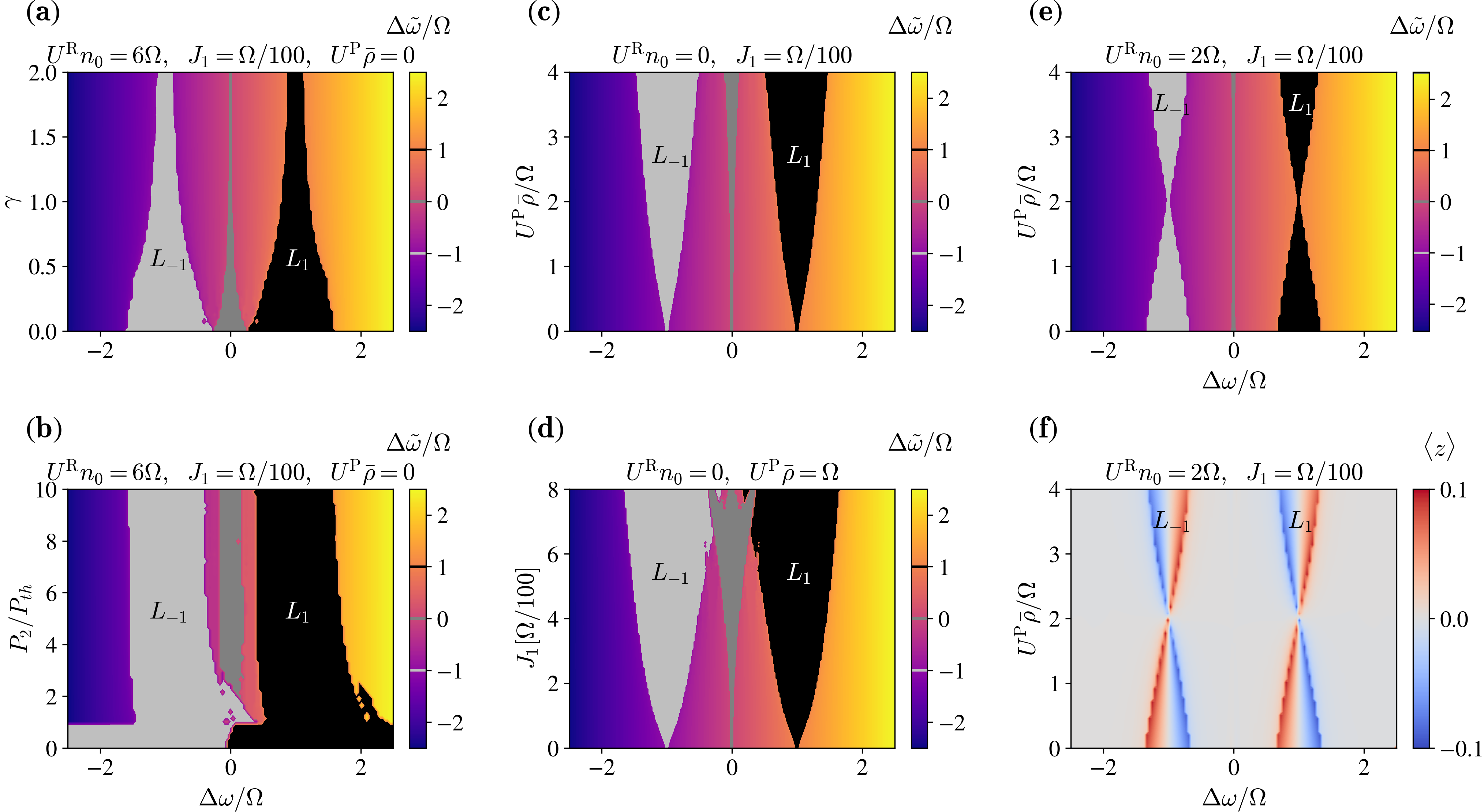}
    \end{center}
    \vspace{-0.5 cm}
\caption{Color maps of the obtained dressed detuning functionality (with main integer-locked regions highlighted in gray scale) versus bare detuning $\Delta\omega$ and as a function of: (a) the polariton decay rate $\gamma$ for $P_1/P_{th}=P_2/P_{th}=10$ for $U^\mathrm{R}n_0=6\Omega$; (b) the degree of condensation of mode $2$, $P_2/P_{th}$, while mode one is in the large condensation limit and $U^\mathrm{R}n_0=6\Omega$;  (c) [(e)] the polariton-polariton interaction $U^\mathrm{P}\bar{\rho}$ for $U^\mathrm{R}n_0=0$ [$U^\mathrm{R}n_0=2\Omega$]; and (d) the amplitude of the coupling $J_1$ with $U^\mathrm{P}\bar{\rho}=\Omega$. (f) Map of the time-average of $z(t)$ for the case (e), evidencing the sign dependence of $\langle z\rangle$ with the interactions inside the locking tongues.}
\label{fig:Linear_b}
\end{figure*}

Here, by numerically solving the model of Eq.~\eqref{EQ:GP_ph} as described in Sec.~\ref{sec:fullmodel}, we explore the behavior of the dominant integer locking regions for the linear case of $J(t)= J_1 (e^{i  \Omega t}+e^{-i  \Omega t})$. We choose, without affecting the qualitative conclusions, a $(\phi_0,z_0)$ initial condition with $\phi_0=\pi$ and initial population imbalance $z_0$ consistent with $|\psi^\mathrm{eq}_1|^2$ and $|\psi^\mathrm{eq}_2|^2$, i.e., for the balanced driving case $z_0=0$. Unless otherwise stated the remaining parameters are as given in Fig.~\ref{fig:Basins}.

Figure~\ref{fig:Linear_b} shows the obtained dressed detuning $\Delta\tilde\omega$ as a function of the bare detuning and several other parameters. Figures~\ref{fig:Linear_b}(a) and~\ref{fig:Linear_b}(b) show the locking regions dependence with $\gamma$ (for $P_1=P_2$) and with $P_2/P_{th}$, respectively, for fixed interactions and amplitude of the mechanical-induced coupling, $J_1$. Since we take $P_1/P_{th}=10$ in both cases, we use the gain expression Eq.~\eqref{eq:gammaFull} instead of the strongly condensed limit. As expected from the RWA mapping to the known results of time-independent $J$ given in Sec.~\ref{sec:J0}, the widths of the locking decrease with increasing $\gamma$. Otherwise, these widths remain almost constant when one of the modes is less condensed than the other. We see that even for small $J_1/\Omega$ the interaction makes the asynchronous locking a robust feature, as it is the case of the synchronization discussed in Sec.~\ref{sec:J0} \cite{Wouters2008}. 

On the other hand, Fig.~\ref{fig:Linear_b}(b) shows that for $P_2/P_{th}<1$ the regions with $\ell=-1$ and $1$ instead of forming tongues they occupy the whole bare detuning space. However, in contrast to the $P_2/P_{th}>1$ cases, these locking conditions are not related to locking of two self-sustained oscillators. To understand this notice that, for $P_2/P_{th}<1$, if mode $2$ were isolated its losses would surpass the gains and it would be empty, i.e., it is not a self-sustained oscillator. The amplitude of the strongly populated mode $1$, $\psi_1(t)$---which, if we neglect mode $2$, can be approximated as $\psi_1(t)=|\psi^\mathrm{eq}_1|\mathrm{e}^{-\mathrm{i}\omega_1 t}$ where we are assuming that $\omega_1$ includes the effect of the local interactions---enters to the equation of motion of the empty mode $2$ multiplied by $J(t)$. Mode $2$ then becomes a forced oscillator subject to a two-tones driving of frequencies, $\omega_{\pm}=\omega_1\pm\Omega$. It follows that $\Delta\tilde\omega$ can only be $\ell\Omega$ with $\ell=-1$ or $1$ depending on which tone dominates and becomes $\tilde\omega_2$. As the mode natural frequency is $\omega_2=\Delta\omega+\omega_1$ comparison with $\omega_\pm$ leads to $\mathrm{sign}(\ell)=\mathrm{sign}(\Delta\omega)$.

The dependence of the dressed detuning on $U^{\textrm{P}}\bar\rho$ is shown in Fig.~\ref{fig:Linear_b}(c), for $U^{\textrm{R}}n_0=0$ and $J_1/\Omega=0.01$. The main locking zones get larger with this interaction, which is also in agreement with the discussion in Sec.~\ref{sec:rwa} and the RWA-associated time-independent $J$ results of Sec.~\ref{sec:J0}. The control of $U^{\textrm{P}}\bar\rho$ can be experimentally realized by modifying the power for constant  $U^{\textrm{P}}$---within the strong condensation limit this can be simulated by a direct variation of $\bar\rho$. In Fig.~\ref{fig:Linear_b}(d) we fix the interactions and explore the dependence of $\Delta\tilde\omega$ as a function of $J_1$. This panel also shows growing widths of locking regions with $\ell=\pm 1$ and $0$. At larger values of $J_1$ these zones overlap one another, reflecting that, similar to what is shown in Fig.~\ref{fig:Basins} as a function of the interaction, the system would reach different attractors depending on the initial conditions. We note that for even stronger coupling and/or stronger interactions the main locking regions disappear or coexist with fractional locking states and even chaotic attractors as the ones discussed in Secs.~\ref{sec:frac} and \ref{sec:doub}.

Finally, the effect of the competition between the polariton-polariton and reservoir-polariton interactions can be seen in Fig.~\ref{fig:Linear_b}(e) where $U^{\textrm{P}}\bar\rho$ is modified while $U^{\textrm{R}} n_0\neq0$ is kept constant. For small $U^{\textrm{P}}\bar\rho$ the width of the locking regions is enhanced by the dominance of $U^{\textrm{R}} n_0$ then, as $U^{\textrm{P}}\bar\rho$ grows, the width decreases up to a minimum in which both interactions neutralize each other, then for larger $U^{\textrm{P}}\bar\rho$ the enhancement due to polariton-polariton interaction dominates and the width of the locking region grows again. Figure~\ref{fig:Linear}(f) shows that the sign of $\langle z\rangle$ reverses depending on which interaction prevails. This behavior is consistent with the approximation of the dressed detuning given in Eq.~\eqref{eq:DressedDet} as it explains why the locking region is minimal when both interactions contribute amounts proportional to $\langle z\rangle$ that cancel each other.

\clearpage




\end{document}